\documentclass[journal,12pt,draftclsnofoot,onecolumn,english]{IEEEtran}

\usepackage{epsfig}
\usepackage{times}
\usepackage{float}
\usepackage{afterpage}
\usepackage{amsmath}
\usepackage{amstext,cite}
\usepackage{amssymb,bm}
\usepackage{latexsym}
\usepackage{color}
\usepackage{graphicx}
\usepackage{amsmath}
\usepackage{amsthm}
\usepackage{graphicx}
\usepackage[center]{caption}
\usepackage{pstricks}
\usepackage{caption}
\usepackage{subcaption}
\usepackage{booktabs}
\usepackage{multicol}
\usepackage{lipsum}
\usepackage[T1]{fontenc}
\usepackage{hyperref}
 \usepackage{aecompl}
 \usepackage{mathrsfs}

\usepackage{soul}
\usepackage{cancel}

\usepackage{changes}
\definechangesauthor[color=red,name={Mingyue Ji}]{MJ}

\allowdisplaybreaks

\long\def\comment#1{}

\newcommand{\uv}{{\mathbf u}}

\newcommand{\xv}{{\mathbf x}}
\newcommand{\yv}{{\mathbf y}}

\newcommand{\Ac}{{\mathcal A}}
\newcommand{\Bc}{{\mathcal B}}

\newcommand{\Dc}{{\mathcal D}}

\newcommand{\Fc}{{\mathcal F}}

\newcommand{\Ic}{{\mathcal I}}

\newcommand{\Lc}{{\mathcal L}}

\newcommand{\Nc}{{\mathcal N}}

\newcommand{\Sc}{{\mathcal S}}

\newcommand{\Wc}{{\mathcal W}}
\newcommand{\Vc}{{\mathcal V}}
\newcommand{\Xc}{{\mathcal X}}
\newcommand{\Yc}{{\mathcal Y}}

\newcommand{\qsf}{{\mathsf q}}

\newcommand{\Bsf}{{\mathsf B}}

\newcommand{\Ksf}{{\mathsf K}}

\newcommand{\Msf}{{\mathsf M}}
\newcommand{\Nsf}{{\mathsf N}}

\newcommand{\Rsf}{{\mathsf R}}

\newtheorem{thm}{Theorem}
\newtheorem{cor}{Corollary}
\newtheorem{lem}{Lemma}

\newtheorem{rem}{Remark}

\providecommand{\definitionname}{Definition}

\begin{document}

\title{On Optimal Load-Memory Tradeoff of Cache-Aided Scalar Linear  Function Retrieval} 
\author{
Kai~Wan,~\IEEEmembership{Member,~IEEE,} 
Hua~Sun,~\IEEEmembership{Member,~IEEE,}
Mingyue~Ji,~\IEEEmembership{Member,~IEEE,}  
Daniela Tuninetti,~\IEEEmembership{Senior~Member,~IEEE,}
and~Giuseppe Caire,~\IEEEmembership{Fellow,~IEEE}
\thanks{A short version of this paper was submitted to 
   the IEEE 2020 International Symposium on Information Theory, Los Angeles, California, USA.}
\thanks{
K.~Wan and G.~Caire are with the Electrical Engineering and Computer Science Department, Technische Universit\"at Berlin, 10587 Berlin, Germany (e-mail:  kai.wan@tu-berlin.de; caire@tu-berlin.de). The work of K.~Wan and G.~Caire was partially funded by the European Research Council under the ERC Advanced Grant N. 789190, CARENET.}
\thanks{
H.~Sun is with the Department of Electrical Engineering, University of North Texas, Denton, TX 76203 (email: hua.sun@unt.edu).
}
\thanks{
M.~Ji is with the Electrical and Computer Engineering Department, University of Utah, Salt Lake City, UT 84112, USA (e-mail: mingyue.ji@utah.edu). The work of M.~Ji was supported in part by NSF Awards 1817154 and 1824558.}
\thanks{
D.~Tuninetti is with the Electrical and Computer Engineering Department, University of Illinois at Chicago, Chicago, IL 60607, USA (e-mail: danielat@uic.edu). The work of D.~Tuninetti was supported in part by NSF Award 1910309.}
}
\maketitle

\begin{abstract}
 Coded caching has the potential to greatly reduce network traffic by leveraging the cheap and abundant storage available in end-user devices so as to create multicast opportunities in the delivery phase.  In the seminal work by Maddah-Ali and Niesen (MAN), the shared-link coded caching problem was formulated, where each user demands one file (i.e., single file retrieval). This paper generalizes the MAN problem so as to allow users to request {\it scalar linear functions} of the files. 
This paper proposes a novel coded delivery scheme that, based on MAN uncoded cache placement, is shown to allow for the decoding of arbitrary scalar linear functions of the files (on arbitrary finite fields).  
 Interestingly, and quite surprisingly, it is shown that the load for cache-aided scalar linear function retrieval depends on the number of linearly independent functions that are demanded, akin to the cache-aided single-file retrieval problem where the load depends on the number of distinct file requests. 
The proposed scheme is optimal under the constraint of uncoded cache placement, in terms of worst-case load, and within a factor 2 otherwise. 
%
The key idea of this paper can be 
extended to all scenarios which the original MAN scheme has been extended to, 
including demand-private and/or device-to-device settings.  
\end{abstract}

\begin{IEEEkeywords}
Coded caching; 
uncoded cache placement; 
linear scalar function retrieval.
\end{IEEEkeywords}

\section{Introduction}
\label{sec:intro}

Information theoretic coded caching was originally proposed by Maddah-Ali and Niesen (MAN) in~\cite{dvbt2fundamental} for the shared-link caching systems containing a server with a library of $\Nsf$ equal-length files, which is connected to $\Ksf$ users through a noiseless shared-link. Each user can store $\Msf$ files in their local cache. 
Two phases are included in the MAN  caching system: i) cache placement phase: content is pushed into each cache without knowledge of future demands; ii) delivery phase: each user demands one file,  and according to the cache contents, the server broadcasts coded packets to all the users.
The objective is to minimize the transmitted load (i.e., number of transmitted bits normalized by the length of a single file) to satisfy the all the user' demands.

The MAN coded caching scheme proposed in~\cite{dvbt2fundamental}, uses a combinatorial design in the placement phase (referred to as MAN cache placement), such that in the delivery phase binary multicast messages (referred to as MAN multicast messages) can simultaneously satisfy the demands of users. Under the constraint of uncoded
cache placement (i.e., each user directly caches a subset of the library bits), the MAN scheme can achieve the minimum worst-case load among all possible demands when $\Nsf \geq  \Ksf$~\cite{ontheoptimality}. On the observation that if 
if there are files demanded multiple times, some MAN multicast messages can be obtained as a binary linear combination of other MAN multicast messages, Yu, Maddah-Ali, and Avestimehr (YMA) proposed an improved delivery scheme in~\cite{exactrateuncoded}. The YMA delivery, with MAN placement, achieves the minimum worst-case load  under the constraint of uncoded cache placement. The cost of uncoded cache placement compared to coded cache placement was proved in~\cite{yufactor2TIT2018} to be at most $2$.

MAN coded caching~\cite{dvbt2fundamental} has been extended to numerous models, such as 
Device-to-Device (D2D) caching systems~\cite{d2dcaching}, 
private coded caching~\cite{wan2019privatecaching,wan2019d2dprivate}, 
coded distributed computing~\cite{distributedcomputing}, and
coded data shuffling~\cite{neartoptimalAttia2018,fundamentalshuffling2018,decentralizedDS2019wan} -- just to name a few.
%
A common point of these models is that each user requests one file -- some allow for users to request (the equivalent of) multiple files~\cite{ji2015multirequest,Sengupta2017multirequest,distributedcomputing,neartoptimalAttia2018,fundamentalshuffling2018,decentralizedDS2019wan} which however does not change much the nature of the problem. 
In general, linear and multivariate polynomial operations are widely used fundamental primitives for building the complex queries that support on-line analytics and data mining procedures. For example, linear
operations are critical in modern deep learning and artificial intelligence algorithms, where matrix-matrix or matrix-vector multiplications are at the core of iterative optimization algorithms; while algebraic polynomial
queries naturally arise in engineering problems such as those described by differential equations and distributed machine learning algorithms \cite{so2019codedprivateml,aliasgari2019private}. In those scenarios, it may be too resource-consuming (in terms of bandwidth, or execution time, or storage space)
to download locally all the input variables in order to compute the desired output value. Instead, it is desirable to directly download the result of the desired output function. 
 This paper studies the fundamental tradeoff between local storage and network load when users are interested in retrieving a function of the dataset available at the server. 
 
 
The question we ask in this paper is, compared to the original MAN caching problem,  whether the optimal worst-case load 
is increased  when the users are allowed to request {\it scalar linear functions of the files}  -- the first non-trivial extension of the MAN single-file-retrieval problem, on the way to understand the problem of retrieving general functions. 
The original MAN shared-link caching problem in~\cite{dvbt2fundamental} where each user request one file is a special case of the formulated shared-link cache-aided  scalar linear function retrieval problem. 
 
 In addition to the novel problem formulation,  our main results are as follows:
\begin{itemize}

\item 
{\it Achievable scheme for demanded functions on the binary field.}
We start by considering the case of scalar linear functions on $\mathbb{F}_2$.
Based on the YMA delivery, which uses an ``interference cancellation'' idea on $\mathbb{F}_2$, we propose a novel delivery scheme 
 whose key idea is to deliver only the largest set of linearly independent functions, while the remaining ones can be reconstructed by proper linear combinations of those already retrieved. This can be thought of as the generalization of the idea to only deliver the files requested by the ``leader users'' in the YMA delivery. 

\item 
{\it Generalization to demanded functions on arbitrary finite field.}
We then generalize the proposed scheme to the case where the demands are scalar linear functions on an arbitrary finite field $\mathbb{F}_{\qsf}$. 
 To the best of our knowledge, even for the originally MAN coded caching problem, no caching scheme is known in the literature for arbitrary finite fields.  
Compared to the YMA delivery scheme,  we use  different encoding (based on a finite field interference alignment idea) and decoding procedures that work on an arbitrary finite field.

{\it Interestingly, the achieved load by the proposed scheme only depends on the number of linearly independent functions that are demanded, akin to the YMA's cache-aided single-file retrieval scheme where the load depends on the number of distinct file requests.}

\item 
{\it Optimality.}
On observation that the converse bound for the original MAN caching problem in~\cite{ontheoptimality,exactrateuncoded} is also a converse in the considered cache-aided function retrieval problem, we prove that the proposed scheme achieves the optimal worst-cast load  under the constraint of uncoded cache placement.
Moreover,  the achieved  worst-case load of the proposed scheme is also proved to be order optimal in general within a factor of $2$.

From the results in this paper, we can answer the question we asked at the beginning of this paper: 
{\it the optimal worst-case load  under the constraint of uncoded cache placement is not increased  when the users are allowed to request scalar linear functions of the files.}

\end{itemize}

\subsection{Paper Organization}
The rest of this paper is organized as follows.
Section~\ref{sec:model} formulates the cache-aided function retrieval problem and introduces some related results in the literature.
Section~\ref{sec:main} provides and discusses the main results in this paper. 
Section~\ref{sec:gf2} and Section~\ref{sec:gfq} describe the proposed achievable caching schemes on the binary field and on arbitrary finite field, respectively.
Section~\ref{sec:conclusion} concludes the paper.
Some of the proofs are given in the Appendices.

\subsection{Notation Convention}
Calligraphic symbols denote sets, 
bold symbols denote vectors, 
and sans-serif symbols denote system parameters.
We use $|\cdot|$ to represent the cardinality of a set or the length of a vector;
$[a:b]:=\left\{ a,a+1,\ldots,b\right\}$ and $[n] := [1,2,\ldots,n]$; 
$\oplus$ represents bit-wise XOR; $\mathbb{E}[\cdot]$ represents the expectation value of a random variable; 
$[a]^+:=\max\{a,0\}$; 
$a!=a\times (a-1) \times \ldots \times 1$ represents the factorial of $a$;
$\mathbb{F}_q$ represents a  finite field with order $q$; 
$\text{rank}_q(\mathbb{A})$ represents the rank of matrix $\mathbb{A}$ on field   $\mathbb{F}_q$;
$\text{det}(\mathbb{A})$ represents the determinant matrix $\mathbb{A}$;
$\mathbb{A}_{\Sc,\Vc}$ represents the sub-matrix of $\mathbb{A}$ by selecting from $\mathbb{A}$, the rows with indices in $\Sc$  and the columns with indices in $\Vc$.
we let $\binom{x}{y}=0$ if $x<0$ or $y<0$ or $x<y$.
In this paper, for each set  of integers  $\Sc$, we sort the elements in $\Sc$ in an increasing order and denote the $i^{\text{th}}$ smallest element by $\Sc(i)$, i.e., $\Sc(1)<\ldots<\Sc(|\Sc|)$.

\section{System Model and Related  Results}
\label{sec:model}

\subsection{System Model}
\label{sub:system}
A $(\Ksf,\Nsf,\Msf,\qsf)$ shared-link cache-aided scalar linear function retrieval problem  is defined as follows. 
A central server has access to a library of $\Nsf$ files.
The files are denoted as $F_1,\ldots, F_{\Nsf}$. 
Each file has $\Bsf$ independent and uniformly distributed symbols over a finite field $\mathbb{F}_{\qsf}$, for some prime-power $\qsf$.
The central server is connected to $\Ksf$ users through an error-free 
shared-link. 
Each user is equipped with a cache that can store up to $\Msf \Bsf$ symbols, where $\Msf \in [0,\Nsf]$.

The system operates in two phases.

{\it Cache Placement Phase.} 
During the cache placement phase, each user stores information about the $\Nsf$ files in its local cache without knowledge of future users' demands,
that is, there exist placement functions  $\phi_k, \ k\in[\Ksf]$, such that
\begin{align}
\phi_k &: [\mathbb{F}_{\qsf}]^{\Bsf\Nsf} \to [\mathbb{F}_{\qsf}]^{\Bsf\Msf},  
\label{eq: placement functions def}
\end{align}
We denote the content in the cache of user $k\in[\Ksf]$ by $Z_{k}=\phi_k(F_1, \ldots, F_\Nsf)$. 

{\it Delivery Phase.} 
During the delivery phase, each user requests one scalar linear function of the files. 
The demand of user $k \in [\Ksf]$ is represented by the row vector $\yv_k=(y_{k,1}, \ldots, y_{k,\Nsf}) \in [\mathbb{F}_{\qsf}]^{\Nsf}$,
which means that user $k$ wants to retrieve $y_{k,1} F_1 +\ldots+ y_{k,\Nsf}  F_{\Nsf}$. 
%
We denote the demand matrix of all users by
\begin{align}
\mathbb{D}=[\yv_1;\ldots;\yv_{\Ksf}] \in [\mathbb{F}_{\qsf}]^{\Ksf \Nsf}.
\label{eq:def of demand matrix}
\end{align} 
 
Given the demand matrix $\mathbb{D}$, the server broadcasts the message $X = \psi(\mathbb{D}, F_1, \ldots, F_\Nsf)$ 
to  each user $k\in [\Ksf]$, where the encoding function $\psi$ is such that
\begin{align}
\psi &: [\mathbb{F}_{\qsf}]^{\Ksf\Nsf} \times [\mathbb{F}_{\qsf}]^{\Bsf\Nsf} \to [\mathbb{F}_{\qsf}]^{\Bsf\Rsf},  
\label{eq: encoding function def}
\end{align}
 for some non-negative $\Rsf$.

{\it Decoding.} 
Each user $k\in [\Ksf]$ decode its desired function from $(\mathbb{D},Z_k,X)$. 
In other words, there exist decoding functions $\xi_k, \ k\in[\Ksf]$, such that
\begin{align}
\xi_k &: [\mathbb{F}_{\qsf}]^{\Ksf\Nsf} \times [\mathbb{F}_{\qsf}]^{\Bsf\Msf} \times [\mathbb{F}_{\qsf}]^{\Bsf\Rsf} \to [\mathbb{F}_{\qsf}]^{\Bsf},  
\\
\xi_k(\mathbb{D},Z_k,X) &=  y_{k,1} F_1 +\ldots+ y_{k,\Nsf}  F_{\Nsf}. 
\label{eq: decoding functions def}
\end{align}

{\it Objective.}
For a given memory size $\Msf \in [0,\Nsf]$, our objective is to determine the {\it minimum worst-case load} among all possible demands,
defined as the smallest $\Rsf$ such that there exist 
placement functions $\phi_k, k\in [\Ksf],$ 
encoding function $\psi$, and 
decoding functions $\xi_k, k\in [\Ksf],$ 
satisfying all the above constraints.
The optimal load is denoted as  $R^{\star}$. 

If each user directly copies  some symbols of the $\Nsf$ files  into its cache, the cache placement is said to be {\it uncoded}. 
The minimum worst-case load under the constraint of uncoded cache placement is denoted by $\Rsf^{\star}_{\text{u}}$.

\subsection{Review of the MAN~\cite{dvbt2fundamental} and YMA~\cite{exactrateuncoded}  Coded Caching Schemes}
\label{sub:review of existing schemes}
In the following, we review the MAN and YMA coded caching schemes, which are on the binary field $\mathbb{F}_2$, for the shared-link caching problem, where each user requests one file. 
 
\paragraph*{MAN Scheme}
{\bf File Split.} Let $t\in [0:\Ksf]$. Partition each file $F_i, i\in [\Nsf]$, into $\binom{\Ksf}{t}$ equal-length subfiles denoted as
\begin{align}
 F_i=\{F_{i,\Wc}:\Wc\subseteq [\Ksf], |\Wc|=t\}.
 \label{eq:MAN filesplit}
\end{align}

{\bf Placement Phase.} User $k\in [\Ksf]$ caches $F_{i,\Wc}, i\in [\Nsf]$, if $k\in \Wc$. 
Hence, each user caches $\Nsf \binom{ \Ksf-1}{t-1}$ subfiles, each of which contains $\frac{\Bsf}{ \binom{\Ksf}{t}}$ symbols, which requires
\begin{align}
 \Msf=\frac{\Nsf t}{\Ksf}.
 \label{eq:MAN memory} 
\end{align}
 
{\bf Delivery Phase.} User $k\in [\Ksf]$ requests the file with index $d_k \in [\Nsf]$. 
The server then broadcasts 
the following {\it MAN multicast messages}: for each $\Sc \subseteq [\Ksf]$ where $|\Sc|=t+1$, the server sends
\begin{align}
 W_{\Sc}= \underset{k\in \Sc}{\oplus } F_{d_k,\Sc\setminus \{k\}}.
 \label{eq:MAN multicastmessage} 
\end{align}  

{\bf Decoding.} The multicast message $W_{\Sc}$ in~\eqref{eq:MAN multicastmessage} is useful to each user $k \in \Sc$, since this user caches all subfiles contained by $W_{\Sc}$ except for the desired subfile $F_{d_k, \Sc \setminus \{k\}}$. 
Considering all multicast messages, each user can recover all uncached subfiles and thus recover its demanded file.

{\bf Load.} The achieved memory-load tradeoff of the MAN   scheme is the lower convex envelop of the following points
\begin{align}
(\Msf, \Rsf)= \left(\frac{\Nsf t}{\Ksf}, \frac{\binom{\Ksf}{t+1}}{\binom{\Ksf}{t}}  \right), \forall t\in [0:\Ksf].  
\label{eq:MAN load}
\end{align}

\paragraph*{YMA Scheme}
{\bf File splitting} and {\bf cache placement} are as for the MAN scheme.

{\bf Delivery Phase.} The main idea of the YMA delivery is that, when a file is demanded by multiple users, some MAN multicast messages in~\eqref{eq:MAN multicastmessage} can be obtained as a linear combinations of others. Thus the load of the MAN scheme in~\eqref{eq:MAN load} can be further reduced by removing the redundant MAN multicast messages. 
More precisely, for each demanded file, randomly choose one user among all users demanding this file and designate it as the ``leader user'' for this file. 
Let $\Dc := \cup_{k\in[\Ksf]} \{d_k\}$ be the set of all distinct files that are demanded, and $\Lc$ be the set of $|\Dc|$ leader users.
The server only sends those multicast message $W_{\Sc}$ in~\eqref{eq:MAN multicastmessage} that are useful for the leader users, that is, if $\Sc \cap \Lc \not=\emptyset$, thus saving $\binom{\Ksf- |\Dc|}{t+1}$ transmissions.

{\bf Decoding.} Clearly, all leaders users can decode their demanded files as per the MAN scheme. The non-leader users appear to miss the multicast message $W_{\Ac}$  for each $\Ac\subseteq [\Ksf]$ where $\Ac\cap \Lc =\emptyset$ and $|\Ac|=t+1$.   It was proved in~\cite{exactrateuncoded} that  
\begin{align}
\underset{\Fc \in \mathscr{F}_{\Bc}}{\oplus } W_{\Bc \setminus \Fc}=0,\label{eq:YMA interference cancelation}
\end{align}
where $\Bc =\Ac\cup \Lc$, and $\mathscr{F}_{\Bc}$ is the family of  subsets  $\Fc \subseteq \Bc$, where   each file in $\Dc$ is requested by exactly one user in $\Fc$.  The key observation is that in $\underset{\Fc \in \mathscr{F}_{\Bc} }{\oplus } W_{\Bc \setminus \Fc}$ each involved subfile appears exactly twice (i.e., contained into two MAN multicast messages)\footnote{\label{foot:appear contain} In this paper, $A$ `appears' in a linear combination means that in the linear combination, there exists some term  in the linear combination including $A$.  A linear combination `contains' $B$ means that in the linear combination, the total coefficient of $B$ is not $0$. For example, we say $A$ appears in the linear combination $(A\oplus B)\oplus (A \oplus C)$, but the linear combination does not contain $A$.}, whose contribution on $\mathbb{F}_2$ is thus zero. 
 From~\eqref{eq:YMA interference cancelation}, we have
\begin{align}
W_{\Ac} =\underset{\Fc \in \mathscr{F}_{\Bc}: \Fc\neq \Lc }{\oplus } W_{\Bc \setminus \Fc}.
\label{eq:YMA reconstruction}
\end{align}
In other words,  the multicast message $W_{\Ac}$ can be reconstructed by all users from the delivery phase.

{\bf Load.}  The YMA scheme requires the load of  
\begin{align}
\frac{\binom{\Ksf}{t+1}-\binom{\Ksf- |\Dc|}{t+1}}{\binom{\Ksf}{t}},\label{eq:YMA load}
\end{align}
if the set of the demanded files is $\Dc$.
The worst-case load is attained for $|\Dc|=\min(\Nsf,\Ksf)$, thus the achieved memory-load tradeoff of the YMA scheme is the lower convex envelop of the following points
\begin{align}
(\Msf, \Rsf)= \left(\frac{\Nsf t}{\Ksf}, \frac{\binom{\Ksf}{t+1}-\binom{\Ksf- \min(\Nsf,\Ksf) }{t+1}}{\binom{\Ksf}{t}}  \right), \forall t\in [0:\Ksf]. 
\label{eq:YMA worst load}
\end{align} 

\section{Main Results and Discussion}
\label{sec:main} 
In this section, we summarize the main results in this paper. 
 
The proposed caching scheme in Section~\ref{sec:gf2} (for $\qsf=2$) and Section~\ref{sec:gfq} (for general prime-power $\qsf$),  achieves the following load.
\begin{thm}[Achievability]
\label{thm:achieved load}
For the $(\Ksf,\Nsf,\Msf,\qsf)$ shared-link cache-aided scalar linear function retrieval problem, the YMA load in~\eqref{eq:YMA worst load} is an achievable worst-case load. 
More precisely, for cache size $\Msf = \frac{\Nsf t}{\Ksf}$, with $t\in[0:\Ksf]$, and for demand matrix $\mathbb{D}$, the load 
\begin{align}
\Rsf(\mathbb{D}) := \frac{\binom{\Ksf}{t+1}-\binom{\Ksf-\text{rank}_{\qsf} (\mathbb{D}) }{t+1}}{ \binom{\Ksf}{t}} 
\label{eq:achieved load}
\end{align} 
is achievable.  The worst-case load is attained by $\text{rank}_{\qsf} (\mathbb{D}) =\min(\Nsf,\Ksf)$. 
\hfill $\square$ 
\end{thm}

\begin{rem}[Dependance on the rank of the demand matrix] 
\label{rem:Comparison to YMA}
\rm
The load in~\eqref{eq:achieved load} is a generalization of the load in~\eqref{eq:YMA load} achieved by the YMA scheme. 
More precisely, if each user $k\in [\Ksf]$ requests one file  (i.e., $\yv_k\in[0:1]^{\Nsf}$ with a unit norm),  $\text{rank}_{\qsf} (\mathbb{D})$ is exactly the number of demanded files, and thus  the proposed scheme  achieves the   load in~\eqref{eq:YMA load} as the YMA scheme.
Interestingly, the load of the proposed scheme only depends on the rank of the demand matrix of all users,  instead of on the specifically demanded functions. 
\hfill $\square$ 
\end{rem}

\begin{rem}[High-level ideas to derive the load in Theorem~\ref{thm:achieved load}]
\label{rem:high level ideas}
\rm
We partition the ``symbol positions" set $[\Bsf]$ as follows
\begin{align}
[\Bsf] = \{ \Ic_\Wc : \Wc\subseteq [\Ksf], |\Wc|=t \} \ \text{such that} \ |\Ic_\Wc| = \Bsf/\binom{\Ksf}{t}.
\label{eq:position partition}
\end{align}
Then, with a Matlab-inspired notation, we let
\begin{align}
F_{i,\Wc} = F_i(\Ic_\Wc), \ \forall  \Wc\subseteq [\Ksf] : |\Wc|=t, \ \forall i\in[\Nsf],
\label{eq:new file split}
\end{align}
representing the set of symbols of $F_i$ whose position is in $\Ic_\Wc$.
As in the MAN placement, user $k\in[\Ksf]$ caches $F_{i,\Wc}$ if $k\in\Wc$.
By doing so, any scalar linear function is naturally partitioned into ``blocks'' as follows
\begin{align}
y_{k,1} F_1 +\ldots+ y_{k,\Nsf}  F_{\Nsf}
&=
\{\underbrace{y_{k,1} F_1(\Ic_\Wc) +\ldots+ y_{k,\Nsf}  F_{\Nsf}(\Ic_\Wc)}_{\text{$:=B_{k,\Wc}$ is the $\Wc$-th block of the $k$-th demanded function}}
: \Wc\subseteq [\Ksf], |\Wc|=t
\}.
\label{eq:demand partition}
\end{align} 
Some blocks of the demanded functions can thus be computed based on the cache content available at each user while the remaining ones need to be delivered by the server.
With this specific file split (and corresponding MAN cache placement), {\bf we   operate the MAN  delivery  scheme 
over the blocks instead of over the subfiles}; more precisely, instead of~\eqref{eq:MAN multicastmessage} we transmit 
\begin{align}
  W_{\Sc}= \sum_{k\in \Sc} \alpha_{\Sc,k} B_{k,\Sc\setminus \{k\}}, 
  \forall \Sc \subseteq [\Ksf] : |\Sc|=t+1,  
  \label{eq:function retrieval delivery on Fq} 
\end{align}
for some $\alpha_{\Sc,k} \in \mathbb{F}_{\qsf} \setminus \{0\}$ and where $B_{k,\Wc}$ was defined in~\eqref{eq:demand partition}.
Clearly, this scheme achieves the same load as in~\eqref{eq:MAN load} (and works on any finite field and any $\alpha_{\Sc,k} \in \mathbb{F}_{\qsf} \setminus \{0\}$).

The questions is, whether with~\eqref{eq:function retrieval delivery on Fq}  we can do something similar to the YMA delivery scheme. More specifically, 
\begin{enumerate}
\item what is a suitable definition of the leader user set $\Lc$;
\item what is a suitable choice of $\alpha_{\Sc,k}$'s in~\eqref{eq:function retrieval delivery on Fq};
and 	
\item assuming we only send the multicast messages in~\eqref{eq:function retrieval delivery on Fq} that are useful for the leader users (i.e., $  W_{\Sc}$ where $\Sc \subseteq [\Ksf]$, $|\Sc|=t+1$, and $\Sc \cap \Lc \neq \emptyset$), 
what is the counterpart of~\eqref{eq:YMA reconstruction}; here for each $\Ac \subseteq [\Ksf]$ where $|\Ac|=t+1$ and $\Ac \cap \Lc =\emptyset$,  we seek  
\begin{align}
W_{\Ac}= \sum_{\Sc\subseteq [\Ksf]:|\Sc|=t+1, \Sc\cap \Lc \neq \emptyset}  \beta_{\Ac,\Sc} W_{\Sc}.
\label{eq:function retrieval decoding on Fq}
\end{align}

\end{enumerate}
The novelty of our scheme lays in the answers to these questions as follows:
\begin{enumerate}
\item we first choose $\text{rank}_{\qsf} (\mathbb{D})$ leaders (the set of leader users is denoted by $\Lc$), where the demand matrix of the leaders is full-rank.
\item  When $\qsf=2$ (i.e., on the binary field),  
   lets  $\alpha_{\Sc,k} =1$. When
  $\qsf$ is a prime-power, the proposed scheme in Section~\ref{sec:gfq} separates the demanded blocks by the leaders and non-leaders in $W_{\Sc}$ in~\eqref{eq:function retrieval delivery on Fq} as 
\begin{align}
\sum_{k\in \Sc} \alpha_{\Sc,k} B_{k,\Sc\setminus \{k\}} = \sum_{k_1\in \Sc \cap \Lc}   \alpha_{\Sc,k_1} B_{k_1,\Sc\setminus \{k_1\}} +\sum_{k_2\in \Sc\setminus \Lc}  \alpha_{\Sc,k_2} B_{k_2,\Sc\setminus \{k_2\}};\label{eq:general separation}
\end{align}
we then alternate the coefficients of the desired blocks by the leaders (i.e., users in ${\Sc \cap \Lc}$) between $+1$ and $-1$,
  i.e., the coefficient of the desired block of the first leader
is $+1$, the coefficient of the desired block of the second leader
is $-1$, the coefficient of the desired block of the third leader
is $+1$, etc;
similarly, we alternate the coefficients of the desired blocks by the non-leaders (i.e., users in ${\Sc\setminus \Lc}$) between $+1$ and $-1$.\footnote{\label{foot:from PFR}
This type of code was originally proposed in~\cite{sun2017pc} for the private function retrieval problem, where there is a memory-less user aiming to retrieval a scalar linear function of the files in the library from multiple servers (each server can access to the whole library), while preserving the demand of this user from each server.}
\item With the above encoding scheme, we can compute the decoding coefficients (as $ \beta_{\Ac,\Sc}$ in~\eqref{eq:function retrieval decoding on Fq}) such that~\eqref{eq:function retrieval decoding on Fq} holds for each  $\Ac \subseteq [\Ksf]$ where $|\Ac|=t+1$ and $\Ac \cap \Lc =\emptyset$.
 In other words, each user can recover all multicast messages $W_{\Sc}$ where $\Sc \subseteq [\Ksf]$ and $|\Sc|=t+1$, and thus it can recover its desired function.
\end{enumerate}
\hfill $\square$ 
\end{rem}

Since the setting where each user demands one file is a special case of the considered cache-aided scalar linear function retrieval problem, the converse bounds in~\cite{ontheoptimality,exactrateuncoded,yufactor2TIT2018} for the original shared-link coded caching problem is also a converse in our considered  problem, thus we have:
\begin{thm}[Optimality]
\label{thm:optimal load}
For the $(\Ksf,\Nsf,\Msf,\qsf)$ shared-link cache-aided scalar linear function retrieval problem, under the constraint of uncoded cache placement, the optimal worst-case load-memory tradeoff  is the lower convex envelop of
\begin{align}
(\Msf, \Rsf^{\star}_{\text{u}}) = \left( \frac{\Nsf t}{\Ksf}, \frac{\binom{\Ksf}{t+1}-\binom{\Ksf-\min\{\Ksf,\Nsf\}}{t+1}}{ \binom{\Ksf}{t}} \right), \ \forall t\in [0:\Ksf]. \label{eq:optimal worst case load}
\end{align} 
%
%
%
%
%
Moreover, the achieved worst-case load in~\eqref{eq:optimal worst case load} is optimal within a factor of $2$ in general.
\hfill $\square$ 
\end{thm}

\begin{rem}[Extensions]
\label{rem:extension to other problems}
\rm
  We discuss three extensions of the proposed caching scheme in Theorem~\ref{thm:achieved load} in the following. 

{\it Optimal average load under uncoded and symmetric cache placement.}
We define {\it uncoded and symmetric} cache placement as follows, which is a generalization of   file split in~\eqref{eq:position partition}-\eqref{eq:new file split}.
We partition the ``symbol positions" set $[\Bsf]$ as  
\begin{align}
[\Bsf] = \{ \Ic_\Wc : \Wc\subseteq [\Ksf] \},
\label{eq:general position partition}
\end{align}
and let  $F_{i,\Wc} = F_i(\Ic_\Wc)$ as in~\eqref{eq:new file split}. Each user $k\in[\Ksf]$ caches $F_{i,\Wc}$ if $k\in\Wc$.

Hence, in the delivery phase, user $k$ needs to recover $B_{k,\Wc}$ (defined in~\eqref{eq:demand partition}) where $\Wc \subseteq [\Ksf]\setminus \{k\}$. 
%
%
By  directly using~\cite[Lemma 2]{yufactor2TIT2018} in  the  caching converse bound under uncoded cache placement in~\cite{ontheoptimality,exactrateuncoded}, we can prove that the proposed caching scheme in Theorem~\ref{thm:achieved load} achieves the  minimum average load over  uniform demand distribution under the constraint of uncoded  and symmetric cache placement cross files. 
\begin{cor}
\label{cor:average load}[Optimal average load]
For the $(\Ksf,\Nsf,\Msf,\qsf)$ shared-link cache-aided scalar linear function retrieval problem, under the constraint of uncoded and symmetric cache placement, 
the minimum average load over   uniform demand distribution is the lower convex envelop of 
\begin{align}
(\Msf, \Rsf) = \left( \frac{\Nsf t}{\Ksf}, \mathbb{E}_{\mathbb{D}}\left[\frac{\binom{\Ksf}{t+1}-\binom{\Ksf-\text{rank}_{\qsf} (\mathbb{D}) }{t+1}}{ \binom{\Ksf}{t}}\right] \right), \ \forall t\in [0:\Ksf]. 
\label{eq:optimal average load}
\end{align} 
\end{cor}
 Notice that an  uncoded and asymmetric cache placement can be treated as a special case of the inter-file coded cache placement in the originally MAN caching problem. It is one of the on-going works to derive the converse bound under the constraints of uncoded cache placement for the considered cache-aided function retrieval problem.

{\it Device-to-Device (D2D) cache-aided scalar linear function retrieval. }
Coded caching  was originally used in Device-to-Device networks in~\cite{d2dcaching}, where in the delivery phase  each user broadcasts   packets as functions of its cached content and the users' demands,  to all other users. The authors in~\cite{ourd2dpaper} extended the YMA scheme to D2D networks by dividing the D2D networks into $\Ksf$ shared-link networks, and used the YMA scheme in each shared-link network. 
 Hence,  when users request scalar linear functions, we can use the same method as in~\cite{ourd2dpaper} to  divide the D2D networks into $\Ksf$ shared-link networks, and then use the proposed caching scheme in  Theorem~\ref{thm:achieved load} in each shared-link network.  
\begin{cor}
\label{cor:d2d woestcase load}[D2D cache-aided  scalar linear function retrieval]
For the $(\Ksf,\Nsf,\Msf,\qsf)$ D2D cache-aided scalar linear function retrieval problem, 
the minimum worse-case load is upper bounded by  the lower convex envelop of 
\begin{align}
(\Msf, \Rsf) = \left( \frac{\Nsf t}{\Ksf}, \max_{\mathbb{D}} \frac{\binom{\Ksf-1}{t}- \frac{1}{\Ksf} \sum_{k\in [\Ksf]}  \binom{\Ksf-1-\text{rank}_{\qsf} (\mathbb{D}_{[\Ksf]\setminus \{k\}}) }{t}}{ \binom{\Ksf-1}{t-1}}  \right), \ \forall t\in [\Ksf]. 
\label{eq:d2d woestcase load}
\end{align} 
\end{cor}

{\it Cache-aided private scalar linear function retrieval.}
For the successful decoding of the proposed scheme in  Theorem~\ref{thm:achieved load},  users need to be aware of the demands of other users,  which is not   private. To preserve the privacy of the   demand of each user against other users, we can generate virtual users as in~\cite{wan2019privatecaching}, such that  each of all possible demanded   functions (the total number of possible demanded functions is $\Nsf^{\prime}:=\frac{\qsf^{\Nsf}-1}{\qsf -1 }$) is demanded exactly $\Ksf$ times.  Thus there are totally $\Nsf^{\prime} \Ksf$ real or virtual users in the system.  Then the proposed scheme in  Theorem~\ref{thm:achieved load} can be used to satisfy the demands of all real or virtual users. Since each user cannot distinguish other real users from virtual users, the resulting scheme does not leak any information on the demands of real users.  
\begin{cor}
\label{cor:private woestcase load}[Cache-aided private scalar linear function retrieval]
For the $(\Ksf,\Nsf,\Msf,\qsf)$ D2D cache-aided scalar linear function retrieval problem, 
the minimum   load is upper bounded by  the lower convex envelop of 
\begin{align}
(\Msf, \Rsf) =  \left(\frac{t}{\Nsf^{\prime} \Ksf }\Nsf,
 \frac{\binom{\Nsf^{\prime} \Ksf }{t+1}-\binom{\Nsf^{\prime} \Ksf - \Nsf }{t+1} }{ \binom{\Nsf^{\prime} \Ksf}{t} } \right) , \ \forall t\in [\Nsf^{\prime} \Ksf]. 
\label{eq:optimal private woestcase load}
\end{align} 
\end{cor}
\hfill $\square$ 
\end{rem}

\section{Achievable Scheme in Theorem~\ref{thm:achieved load} for $\qsf=2$} 
\label{sec:gf2} 
In the following, we describe the proposed scheme 
when the demands are scalar linear functions on   $\mathbb{F}_2$.
We start with the following example.

\subsection{Example}
\label{sub:example F2}
Consider the $(\Ksf,\Nsf,\Msf,\qsf)=(6,3,1,2)$ shared-link cache-aided scalar linear function retrieval problem, where $t=\Ksf\Msf/\Nsf=2 $. In the cache placement, each file is partitioned into $\binom{\Ksf}{t}=15$ equal-length subfiles. 
We use the file split in~\eqref{eq:position partition}-\eqref{eq:new file split}, resulting in the demand split in~\eqref{eq:demand partition}.

In the delivery phase, we assume that
\begin{align*}
&\text{user $1$ demands $F_1$;}\\
&\text{user $2$ demands $F_2$;}\\
&\text{user $3$ demands $F_3$;}\\
&\text{user $4$ demands $F_1 \oplus F_2$;}\\
&\text{user $5$ demands $F_1 \oplus F_3$;}\\
&\text{user $6$ demands $F_1 \oplus F_2 \oplus F_3$;}
\end{align*}
i.e., the demand matrix is 
\begin{align}
\mathbb{D}=\left[\begin{array}{ccc}
1 & 0 & 0\\
0 & 1 & 0\\
0 & 0 & 1\\
1 & 1 & 0\\
1 & 0 & 1\\
1 & 1 & 1
\end{array}\right].\label{eq:demand matrix example 1}
\end{align}
On the observation that  $\text{rank}_2(\mathbb{D})=3$, we choose $3$ users as leaders, where the demand matrix of these $3$ leaders is also full-rank. Here, we choose $\Lc=[3]$.

{\it Encoding.}
For each set $\Sc \subseteq [\Ksf]$ where $|\Sc|=t+1=3$, 
we generate a multicast message with $\alpha_{\Sc,k} =1$ in~\eqref{eq:function retrieval delivery on Fq}.
Hence, we have
\begin{subequations}
\begin{align}
& W_{\{1,2,3\}}= F_{1,\{2,3\}} \oplus F_{2,\{1,3\}} \oplus F_{3,\{1,2\}}; \label{eq:W123 F2}\\
&W_{\{1,2,4\}}= F_{1,\{2,4\}} \oplus F_{2,\{1,4\}} \oplus (F_{1,\{1,2\}}\oplus F_{2,\{1,2\}}); \label{eq:W124 F2}\\
& W_{\{1,2,5\}}= F_{1,\{2,5\}} \oplus F_{2,\{1,5\}} \oplus (F_{1,\{1,2\}}\oplus F_{3,\{1,2\}}); \label{eq:W125 F2}\\
& W_{\{1,2,6\}}= F_{1,\{2,6\}} \oplus F_{2,\{1,6\}} \oplus (F_{1,\{1,2\}}\oplus F_{2,\{1,2\}}\oplus F_{3,\{1,2\}}); \label{eq:W126 F2}\\
& W_{\{1,3,4\}}= F_{1,\{3,4\}} \oplus F_{3,\{1,4\}} \oplus (F_{1,\{1,3\}}\oplus F_{2,\{1,3\}}); \label{eq:W134 F2}\\
& W_{\{1,3,5\}}= F_{1,\{3,5\}} \oplus F_{3,\{1,5\}} \oplus (F_{1,\{1,3\}}\oplus F_{3,\{1,3\}}); \label{eq:W135 F2}\\
& W_{\{1,3,6\}}= F_{1,\{3,6\}} \oplus F_{3,\{1,6\}} \oplus (F_{1,\{1,3\}}\oplus F_{2,\{1,3\}}\oplus F_{3,\{1,3\}}); \label{eq:W136 F2}\\
& W_{\{1,4,5\}}= F_{1,\{4,5\}} \oplus (F_{1,\{1,5\}}\oplus F_{2,\{1,5\}}) \oplus (F_{1,\{1,4\}}\oplus F_{3,\{1,4\}}); \label{eq:W145 F2}\\
&W_{\{1,4,6\}}= F_{1,\{4,6\}} \oplus (F_{1,\{1,6\}}\oplus F_{2,\{1,6\}}) \oplus (F_{1,\{1,4\}}\oplus F_{2,\{1,4\}} \oplus F_{3,\{1,4\}}); \label{eq:W146 F2}\\
& W_{\{1,5,6\}}=F_{1,\{5,6\}} \oplus (F_{1,\{1,6\}} \oplus F_{3,\{1,6\}}) \oplus (F_{1,\{1,5\}} \oplus F_{2,\{1,5\}} \oplus F_{3,\{1,5\}}); \label{eq:W156 F2}\\
&W_{\{2,3,4\}}= F_{2,\{3,4\}} \oplus F_{3,\{2,4\}} \oplus (F_{1,\{2,3\}} \oplus F_{2,\{2,3\}}); \label{eq:W234 F2}\\
& W_{\{2,3,5\}}= F_{2,\{3,5\}} \oplus F_{3,\{2,5\}} \oplus (F_{1,\{2,3\}} \oplus F_{3,\{2,3\}}); \label{eq:W235 F2}\\
& W_{\{2,3,6\}}= F_{2,\{3,6\}} \oplus F_{3,\{2,6\}} \oplus (F_{1,\{2,3\}} \oplus F_{2,\{2,3\}} \oplus F_{3,\{2,3\}}); \label{eq:W236 F2}\\
&W_{\{2,4,5\}}= F_{2,\{4,5\}} \oplus (F_{1,\{2,5\}}\oplus F_{2,\{2,5\}}) \oplus (F_{1,\{2,4\}} \oplus F_{3,\{2,4\}}); \label{eq:W245 F2}\\
& W_{\{2,4,6\}}= F_{2,\{4,6\}} \oplus (F_{1,\{2,6\}}\oplus F_{2,\{2,6\}}) \oplus (F_{1,\{2,4\}} \oplus F_{2,\{2,4\}} \oplus F_{3,\{2,4\}}); \label{eq:W246 F2}\\
& W_{\{2,5,6\}}= F_{2,\{5,6\}} \oplus (F_{1,\{2,6\}}\oplus F_{3,\{2,6\}}) \oplus (F_{1,\{2,5\}} \oplus F_{2,\{2,5\}} \oplus F_{3,\{2,5\}}); \label{eq:W256 F2}\\
& W_{\{3,4,5\}}= F_{3,\{4,5\}} \oplus (F_{1,\{3,5\}}\oplus F_{2,\{3,5\}}) \oplus (F_{1,\{3,4\}} \oplus F_{3,\{3,4\}}); \label{eq:W345 F2}\\
&W_{\{3,4,6\}}= F_{3,\{4,6\}} \oplus (F_{1,\{3,6\}}\oplus F_{2,\{3,6\}}) \oplus (F_{1,\{3,4\}} \oplus F_{2,\{3,4\}} \oplus F_{3,\{3,4\}}); \label{eq:W346 F2}\\
& W_{\{3,5,6\}}= F_{3,\{5,6\}} \oplus (F_{1,\{3,6\}}\oplus F_{3,\{3,6\}}) \oplus (F_{1,\{3,5\}} \oplus F_{2,\{3,5\}} \oplus F_{3,\{3,5\}}); \label{eq:W356 F2}\\
&W_{\{4,5,6\}}= (F_{1,\{5,6\}} \oplus F_{2,\{5,6\}}) \oplus (F_{1,\{4,6\}}\oplus F_{3,\{4,6\}}) \oplus (F_{1,\{4,5\}} \oplus F_{2,\{4,5\}} \oplus F_{3,\{4,5\}}). \label{eq:W456 F2}
\end{align}
\label{eq:F2 example WS}
\end{subequations}

{\it Delivery.}
The server broadcasts $W_{\Sc}$ for each $\Sc \subseteq [\Ksf]$ where $|\Sc|=t+1=3$ and $\Sc \cap \Lc \neq \emptyset$. In other words, the server broadcasts all the multicast messages in~\eqref{eq:F2 example WS} except for $W_{\{4,5,6\}}$.

{\it Decoding.}
We show that the untransmitted multicast message $W_{\{4,5,6\}}$ can be reconstructed by the transmitted multicast messages. 
For each set of  users $\Bc \subseteq [\Ksf]$, we  define $\mathscr{V}_{\Bc}$ as the family of  subsets  $\Vc \subseteq \Bc$, where   $|\Vc|=|\Lc| $ and
$\text{rank}_2(\mathbb{D}_{\Vc})=|\Lc|.$
It can be seen that  $\mathscr{V}_{\Bc}$ is the generalization of  $\mathscr{F}_{\Bc}$  defined in the YMA scheme  described in Section~\ref{sub:review of existing schemes}.
When $\Bc=\Lc \cup \{4,5,6\}=[6]$, we have 
\begin{align}
 \mathscr{V}_{[6]} = \Big\{
 &\{1,2,3\},\{1,2,5\},\{1,2,6\},\{1,3,4\},\{1,3,6\},\{1,4,5\},\{1,4,6\},\{1,5,6\},\nonumber\\ 
 &\{2,3,4\},\{2,3,5\},\{2,3,6\},\{2,4,5\},\{2,4,6\},\{3,4,5\},\{3,5,6\},\{4,5,6\}  \Big\}.
 \label{eq:V[6]}
\end{align}

From the above definition, we focus on the following sum of multicast messages  
 \begin{align}
 &\underset{\Vc \in \mathscr{V}_{[6]}  }{\oplus } W_{[6] \setminus \Vc}=0,
  \label{eq:F2 sum including A example}
 \end{align}
where~\eqref{eq:F2 sum including A example} is because  on the LHS of~\eqref{eq:F2 sum including A example}, among all subfiles $F_{i,\Wc}$ where $i\in [3]$, $\Wc\subseteq [6]$, and $|\Wc|=2$, the coefficient of each of $F_{2,\{2,4\}}$, $F_{2,\{2,6\}}$, $F_{2,\{4,6\}}$, $F_{3,\{3,5\}}$, $F_{3,\{3,6\}}$, $F_{3,\{5,6\}}$ is $0$, 
  $F_{1,\{2,3\}}$  appears $4$ times and other subfiles appear $2$ times. 
Hence, the sum is equivalent to $0$ on $\mathbb{F}_2$.
 Notice that in the YMA delivery scheme, the coefficient of each subfile appearing in the sum 
 $\underset{\Fc \in \mathscr{F}_{\Bc} }{\oplus } W_{\Bc \setminus \Fc}$ is $2$. 

We can write~\eqref{eq:F2 sum including A example} as 
\begin{align}
W_{\{4,5,6\}}=\underset{\Vc \in \mathscr{V}_{[6]} : \Vc \neq \Lc }{\oplus } W_{[6] \setminus \Vc}.
\end{align}
In other words, the untransmitted multicast message $W_{\{4,5,6\}}$ can be reconstructed by the transmitted multicast messages. Thus each user can recover all the multicast messages in~\eqref{eq:F2 example WS}, and then recover its desired function.

 {\it Performance.}
In total we transmit  $\binom{\Ksf}{t+1}-\binom{\Ksf-\text{rank}_2 (\mathbb{D}) }{t+1}=\binom{6}{3}-\binom{3}{3}=19$ multicast messages, each of which contains $\frac{\Bsf}{20}$ bits. Hence, the transmitted load is
 $\frac{19}{20}$, which coincides with the optimal worst-case load in Theorem~\ref{thm:optimal load}.

\subsection{General Description}
\label{sub:general F2}
We use the file split in~\eqref{eq:position partition}-\eqref{eq:new file split}, resulting in the demand split in~\eqref{eq:demand partition}.

In the delivery phase, the demand matrix $\mathbb{D}$ is revealed where each element in  $\mathbb{D}$ is either $0$ or $1$.
Among the $\Ksf $ users we first choose $\text{rank}_2(\mathbb{D})$ leaders (assume the set of leaders is $\Lc=\{\Lc(1),\ldots,\Lc(|\Lc|)\}$), where 
\begin{align}
 |\Lc|= \text{rank}_2(\mathbb{D}_{\Lc})= \text{rank}_2(\mathbb{D} ).\label{eq:full rank f2}
\end{align}

{\it Encoding.}
We focus on  each set $\Sc \subseteq [\Ksf]$ where $|\Sc|=t+1$, and generate the multicast message in~\eqref{eq:function retrieval delivery on Fq}  with $\alpha_{\Sc,k} =1$.

{\it Delivery.}
The server broadcasts $W_{\Sc}$ for each $\Sc \subseteq [\Ksf]$ where $|\Sc|=t+1$ and $\Sc \cap \Lc \neq \emptyset$.

{\it Decoding.}
For each set of  users $\Bc \subseteq [\Ksf]$, recall that $\mathscr{V}_{\Bc}$ is the family of  subsets  $\Vc \subseteq \Bc$, where   $|\Vc|=|\Lc|$ and  $ \text{rank}_2(\mathbb{D}_{\Vc})=|\Lc|.$
 We now consider each set $\Ac \subseteq [\Ksf]$ where $|\Ac|=t+1$ and $\Ac\cap \Lc=\emptyset$, and focus on the binary sum
 \begin{align}
 \underset{\Vc \in \mathscr{V}_{\Bc} }{\oplus } W_{\Bc \setminus \Vc} ,\label{eq:F2 sum including A}
 \end{align}
where $\Bc= \Lc \cup \Ac$. 
  A subfile $F_{i,\Wc}$ appears in the sum~\eqref{eq:F2 sum including A} if and only if  $\Wc \subseteq \Bc$ and there exists some user $k\in \Bc\setminus \Wc$ such that $\text{rank}_2(\mathbb{D}_{ \Bc\setminus (\Wc\cup\{k\}) })= |\Lc|$ (i.e., $\mathbb{D}_{ \Bc\setminus (\Wc\cup\{k\}) }$ is full-rank) and $y_{k,i} \neq 0$.
 We then provide the following Lemma,  proved in Appendix~\ref{proof lem F2 decodability}. 
 \begin{lem}
\label{lem:F2 decodability}
If $F_{i,\Wc}$ appears in the sum~\eqref{eq:F2 sum including A}, the number of multicast messages in the sum which contains  $F_{i,\Wc}$ is even.\footnote{\label{foot:even}  Notice that in the YMA scheme for the original MAN caching problem, each subfile in~\eqref{eq:YMA interference cancelation} is contained by   two   multicast messages  in~\eqref{eq:YMA interference cancelation}. Hence,  Lemma~\ref{lem:F2 decodability} is also a generalization of~\cite[Lemma 1]{exactrateuncoded} for the YMA scheme.}
\hfill $\square$ 
\end{lem}

From Lemma~\ref{lem:F2 decodability}, it can be seen that each subfile in the sum~\eqref{eq:F2 sum including A} appears an even number of   times, and thus the coefficient of this subfile in the sum is $0$, 
which leads to
\begin{align}
 W_{\Ac}= \underset{\Vc \in \mathscr{V}_{\Bc} :\Vc \neq \Lc}{\oplus } W_{\Bc \setminus \Vc}.
\end{align}
In other words, $W_{\Ac}$ can be reconstructed by the transmitted multicast messages.

As a result, each user $k$ can recover each multicast message $W_{\Sc}$ where $\Sc \subseteq [\Ksf]$ and $|\Sc|=t+1$, and thus it can decode its desired function.

{\it Performance.}
In total, we   transmit  $\binom{\Ksf}{t+1}-\binom{\Ksf-\text{rank}_2 (\mathbb{D}) }{t+1}$ multicast messages, each of which contains $\frac{\Bsf}{ \binom{\Ksf}{t}}$ bits. Hence, the transmitted load is
\begin{align}
\frac{\binom{\Ksf}{t+1}-\binom{\Ksf-\text{rank}_2 (\mathbb{D}) }{t+1}}{ \binom{\Ksf}{t}}.\label{eq:load F2for each demand}
\end{align} 
 
For the worst-case demands where $\text{rank}_2 (\mathbb{D})$ is full-rank, we have $\text{rank}_2 (\mathbb{D}) =\min\{\Ksf,\Nsf\}$, and we  achieve the worst-case load in~\eqref{eq:optimal worst case load}.

\section{Achievable Scheme in Theorem~\ref{thm:achieved load} for general prime-power $\qsf$} 
\label{sec:gfq} 
In the following, we generalize the proposed caching scheme in Section~\ref{sec:gf2} to the case where the demands are scalar linear functions on arbitrary finite field $\mathbb{F}_\qsf$. 
All the operations in the proposed scheme are on $\mathbb{F}_\qsf$. 
We again start with an example.
  
 
\subsection{Example}
\label{sub:example Fq}

Consider the $(\Ksf,\Nsf,\Msf,\qsf)=(5,3,3/5,\qsf)$ shared-link cache-aided scalar linear function retrieval problem,  where $\qsf$ is a prime-power. In this case, we have  $t=\Ksf\Msf/\Nsf=1$. Hence, in the cache placement,  each file is partitioned into $\binom{\Ksf}{t}= 5$  equal-length subfiles. 
We use the file split in~\eqref{eq:position partition}-\eqref{eq:new file split}, resulting in the demand split in~\eqref{eq:demand partition}.

In the delivery phase, we assume that
\begin{align*}
&\text{user $1$ demands $F_1$;}\\
&\text{user $2$ demands $F_2$;}\\
&\text{user $3$ demands $F_3$;}\\
&\text{user $4$ demands $y_{4,1}F_1 + y_{4,2} F_2+y_{4,3} F_3$;}\\
&\text{user $5$ demands $y_{5,1}F_1 + y_{5,2} F_2+y_{5,3} F_3$;}  
\end{align*}
i.e., the demand matrix is 
\begin{align}
\mathbb{D}=\left[\begin{array}{ccc}
1 & 0 & 0\\
0 & 1 & 0\\
0 & 0 & 1\\
y_{4,1} & y_{4,2} & y_{4,3}\\
y_{5,1} & y_{5,2} & y_{5,3}
\end{array}\right]\in [\mathbb{F}_{\qsf}]^{5 \times 3}.
\label{eq:demand matrix example 2}
\end{align}
We choose the set of leaders $\Lc=[3]$, since $\text{rank}_{\qsf}(\mathbb{D}_{[3]})=3$.

Each user $k\in [\Ksf]$ should recover each block  $B_{k,\Wc}= y_{k,1} F_{1,\Wc} +y_{k,2} F_{2,\Wc}+ y_{k,3}  F_{3,\Wc}$  in the delivery phase, where $\Wc \in [5]\setminus \{k\}$ and $|\Wc|=1$.

{\it Encoding.}
For each set $\Sc \subseteq [\Ksf]$ where $|\Sc|=t+1=2$, recall that the multicast messages are given in~\eqref{eq:function retrieval delivery on Fq} and we separate it as
\begin{align}
 W_{\Sc}
 &= \sum_{k_1\in \Sc \cap \Lc} \alpha_{\Sc,k_1} B_{k_1,\Sc\setminus \{k_1\}} +\sum_{k_2\in \Sc\setminus \Lc} \alpha_{\Sc,k_2}  B_{k_2,\Sc\setminus \{k_2\}} . \label{eq:example 2 Fq WS separated}
\end{align}
We first   alternate  the coefficients (either $ 1$ or $-1$) of the desired blocks of the leaders in $\Sc$, and then   alternate  the coefficients (either $ 1$ or $-1$) of the desired blocks of the non-leaders in $\Sc$.
For example, if $\Sc=\{1,2\}$, we have 
$W_{\{1,2\}}= F_{1,\{2\}}- F_{2,\{1\}}$; if $\Sc=\{1,4\}$, we have   $W_{\{1,4\}}= F_{1,\{4\}}+(y_{4,1} F_{1,\{1\}} +y_{4,2} F_{2,\{1\}}+ y_{4,3}  F_{3,\{1\}})$; if $\Sc=\{4,5\}$, we have 
$W_{\{4,5\}}=(y_{4,1} F_{1,\{5\}} +y_{4,2} F_{2,\{5\}}+ y_{4,3}  F_{3,\{5\}})-(y_{5,1} F_{1,\{4\}} +y_{5,2} F_{2,\{4\}}+ y_{5,3}  F_{3,\{4\}})$.
With this, we can list all the multicast messages as 
\begin{subequations}
\begin{align}
&W_{\{1,2\}}= F_{1,\{2\}}- F_{2,\{1\}}; \label{eq:W12}\\
&W_{\{1,3\}}= F_{1,\{3\}}- F_{3,\{1\}}; \label{eq:W13}\\
&W_{\{1,4\}}= F_{1,\{4\}}+(y_{4,1} F_{1,\{1\}} +y_{4,2} F_{2,\{1\}}+ y_{4,3}  F_{3,\{1\}}); \label{eq:W14}\\
&W_{\{1,5\}}= F_{1,\{5\}}+(y_{5,1} F_{1,\{1\}} +y_{5,2} F_{2,\{1\}}+ y_{5,3}  F_{3,\{1\}}); \label{eq:W15}\\
&W_{\{2,3\}}= F_{2,\{3\}}- F_{3,\{2\}}; \label{eq:W23}\\
&W_{\{2,4\}}= F_{2,\{4\}}+(y_{4,1} F_{1,\{2\}} +y_{4,2} F_{2,\{2\}}+ y_{4,3}  F_{3,\{2\}}); \label{eq:W24}\\
&W_{\{2,5\}}= F_{2,\{5\}}+(y_{5,1} F_{1,\{2\}} +y_{5,2} F_{2,\{2\}}+ y_{5,3}  F_{3,\{2\}}); \label{eq:W25}\\
&W_{\{3,4\}}= F_{3,\{4\}}+(y_{4,1} F_{1,\{3\}} +y_{4,2} F_{2,\{3\}}+ y_{4,3}  F_{3,\{3\}}); \label{eq:W34}\\
&W_{\{3,5\}}= F_{3,\{5\}}+(y_{5,1} F_{1,\{3\}} +y_{5,2} F_{2,\{3\}}+ y_{5,3}  F_{3,\{3\}}); \label{eq:W35}\\
&W_{\{4,5\}}=(y_{4,1} F_{1,\{5\}} +y_{4,2} F_{2,\{5\}}+ y_{4,3}  F_{3,\{5\}})-(y_{5,1} F_{1,\{4\}} +y_{5,2} F_{2,\{4\}}+ y_{5,3}  F_{3,\{4\}}).\label{eq:W45}
\end{align}
\label{eq:Fq example WS}
\end{subequations}

{\it Delivery.}
The server broadcasts $W_{\Sc}$ for each $\Sc \subseteq [\Ksf]$ where $|\Sc|=t+1=2$ and $\Sc \cap \Lc \neq \emptyset$.
In other words, the server broadcasts all the multicast messages in~\eqref{eq:Fq example WS} except for $W_{\{4,5\}}$.

{\it Decoding.}
We first show the untransmitted multicast message $W_{\{4,5\}}$ can be reconstructed by the transmitted multicast messages.  
More precisely, we aim to choose the decoding coefficients $\beta_{\{4,5\},\Sc} \in \mathbb{F}_{\qsf}$ for each $\Sc\subseteq [\Ksf]$ where $|\Sc|=t+1$ and $\Sc\cap \Lc \neq \emptyset$, such that 
\begin{align}
W_{\{4,5\}}= \sum_{\Sc\subseteq [\Ksf]:|\Sc|=t+1, \Sc\cap \Lc \neq \emptyset}  \beta_{\{4,5\},\Sc} W_{\Sc}.\label{eq:decode W45 example 2}
\end{align}
Since on the RHS of~\eqref{eq:decode W45 example 2}  $F_{1,\{4\}}$ only appears in $W_{\{1,4\}}$ and on the LHS of~\eqref{eq:decode W45 example 2} the coefficient of  $F_{1,\{4\}}$ is $-y_{5,1}$, in order to 
have the same coefficient of  $F_{1,\{4\}}$ on both sides of~\eqref{eq:decode W45 example 2}, 
we let 
\begin{align}
 \beta_{\{4,5\},\{1,4\}}= -y_{5,1}= -\text{det}([y_{5,1}]).\label{eq:B14}
\end{align}
Similarly,  we let 
\begin{align}
 \beta_{\{4,5\},\{2,4\}}= -y_{5,2}=-\text{det}([y_{5,2}]), \label{eq:B24}
\end{align}
such that the coefficients of  $F_{2,\{4\}}$ on both sides of~\eqref{eq:decode W45 example 2} are the same; 
let 
\begin{align}
 \beta_{\{4,5\},\{3,4\}}= -y_{5,3}= -\text{det}([y_{5,3}]), \label{eq:B34}
\end{align}
such that the coefficients of  $F_{3,\{4\}}$ on both sides of~\eqref{eq:decode W45 example 2} are the same; 
let 
\begin{align}
 \beta_{\{4,5\},\{1,5\}}=  y_{4,1}= \text{det}([y_{4,1}]), \label{eq:B15}
\end{align}
such that the coefficients of  $F_{1,\{5\}}$ on both sides of~\eqref{eq:decode W45 example 2} are the same;
let 
\begin{align}
 \beta_{\{4,5\},\{2,5\}}=  y_{4,2}= \text{det}([y_{4,2}]), \label{eq:B25}
\end{align}
such that the coefficients of  $F_{2,\{5\}}$ on both sides of~\eqref{eq:decode W45 example 2} are the same;
let 
\begin{align}
 \beta_{\{4,5\},\{3,5\}}=  y_{4,3}= \text{det}([y_{4,3}]), \label{eq:B35}
\end{align}
such that the coefficients of  $F_{3,\{5\}}$ on both sides of~\eqref{eq:decode W45 example 2} are the same.

Next  we focus $ F_{1,\{1\}}$, which appears in $W_{\{1,4\}}$ and $W_{\{1,5\}}$. Since $ \beta_{\{4,5\},\{1,4\}}=-y_{5,1}$ and $\beta_{\{4,5\},\{1,5\}}=  y_{4,1}$,   the coefficient of  $ F_{1,\{1\}}$ on the RHS of~\eqref{eq:decode W45 example 2} is
\begin{align}
y_{4,1} \beta_{\{4,5\},\{1,4\}} +y_{5,1}\beta_{\{4,5\},\{1,5\}}=0.
\end{align}
Similarly, the coefficient of  $ F_{2,\{2\}}$ on the RHS of~\eqref{eq:decode W45 example 2}, which appears in $ W_{\{2,4\}} $  and $ W_{\{2,5\}} $,  is $0$.
The   coefficient of  $ F_{3,\{3\}}$ on the RHS of~\eqref{eq:decode W45 example 2}, which appears in $ W_{\{3,4\}} $  and $ W_{\{3,5\}} $,  is $0$, 

Now we focus on $ F_{1,\{2\}}$, which appears in $W_{\{1,2\}}$, $W_{\{2,4\}}$, and $W_{\{2,5\}}$. Since $ \beta_{\{4,5\},\{2,4\}}=-y_{5,2}$ and $\beta_{\{4,5\},\{2,5\}}=  y_{4,2}$,
in order to let the coefficient of  $ F_{1,\{2\}}$ on the RHS of~\eqref{eq:decode W45 example 2} be $0$, we let 
\begin{align}
 \beta_{\{4,5\},\{1,2\}}=  y_{4,1}  y_{5,2}-y_{5,1} y_{4,2}=\text{det}([y_{4,1},y_{4,2};y_{5,1},y_{5,2}]).\label{eq:B12}
\end{align}
In addition, $F_{2,\{1\}}$ appears in $W_{\{1,2\}}$, $W_{\{1,4\}}$, and  $W_{\{1,5\}}$. The coefficient of $F_{2,\{1\}}$ on the RHS of~\eqref{eq:decode W45 example 2} is
\begin{align}
-\beta_{\{4,5\},\{1,2\}}+  y_{4,2} \beta_{\{4,5\},\{1,4\}}+ y_{5,2} \beta_{\{4,5\},\{1,5\}}=0.
\end{align}
Similarly, we let 
\begin{align}
 \beta_{\{4,5\},\{1,3\}}=y_{4,1} y_{5,3} -y_{5,1}y_{4,3}= \text{det}([y_{4,1},y_{4,3};y_{5,1},y_{5,3}]),\label{eq:B13}
\end{align}
such that  the coefficients of  $F_{1,\{3\}}$ and  $F_{3,\{1\}}$ on the RHS of~\eqref{eq:decode W45 example 2} are $0$.
We let 
\begin{align}
 \beta_{\{4,5\},\{2,3\}}=  y_{4,2}  y_{5,3}  -  y_{5,2}y_{4,3}= \text{det}([y_{4,2},y_{4,3};y_{5,2},y_{5,3}]),\label{eq:B23}
\end{align}
such that  the coefficients of  $F_{2,\{3\}}$ and  $F_{3,\{2\}}$ on the RHS of~\eqref{eq:decode W45 example 2} are $0$.

With the above choice of decoding coefficients, on the RHS of~\eqref{eq:decode W45 example 2}, the coefficients of all the subfiles which is not contained by  $W_{\{4,5\}}$ are $0$.
In addition, the coefficients of each subfile contained by $W_{\{4,5\}}$ are the same on   both sides of~\eqref{eq:decode W45 example 2}. Thus we prove~\eqref{eq:decode W45 example 2}. In conclusion, each user can recover all multicast messages in~\eqref{eq:Fq example WS}, and then recover its demanded function.

{\it Performance.}
In total we transmit  $\binom{\Ksf}{t+1}-\binom{\Ksf-\text{rank}_{\qsf} (\mathbb{D}) }{t+1}=\binom{5}{2}-\binom{2}{2}=9$ multicast messages, each of which contains $\frac{\Bsf}{5}$ symbols. Hence, the transmitted load is $\frac{9}{5}$, which coincides with the optimal worst-case load in Theorem~\ref{thm:optimal load}.

\subsection{General Description}
\label{sub:general Fq}

We use the file split in~\eqref{eq:position partition}-\eqref{eq:new file split}, resulting in the demand split in~\eqref{eq:demand partition}.

In the delivery phase, after the demand matrix $\mathbb{D}$ is revealed, among the $\Ksf $ users we first choose $\text{rank}_{\qsf}(\mathbb{D})$ leaders (assume the set of leaders is $\Lc=\{\Lc(1),\ldots,\Lc(|\Lc|)\}$), where 
\begin{align}
 |\Lc|= \text{rank}_{\qsf}(\mathbb{D}_{\Lc})= \text{rank}_{\qsf}(\mathbb{D} ).\label{eq:full rank}
\end{align}
For each $i\in [|\Lc|]$, we also define that   the {\it leader index}  of leader $\Lc(i)$ is $i$.

From~\eqref{eq:full rank}, we can represent the demands of non-leaders by the linear combinations of the demands of leaders. More precisely, we define  
\begin{align}
F^{\prime}_i:= y_{\Lc(i),1} F_1 +\ldots+ y_{\Lc(i),\Nsf}  F_{\Nsf}, \ \forall i\in[|\Lc|],   \label{eq:demand transformation}
\end{align}
and represent the demand of each user $k\in [\Ksf]$ by
\begin{align}
y_{k,1} F_1 +\ldots+ y_{k,\Nsf}  F_{\Nsf} = x_{k,1} F^{\prime}_1 +\ldots+ x_{k,|\Lc|}  F^{\prime}_{|\Lc|}.\label{eq:new demand representation}
\end{align}
Clearly, for each leader $\Lc(i)$ where $i\in [|\Lc|]$, $ \xv_{\Lc(i)}$ is an $|\Lc|$-dimension unit vector where the $i^{\text{th}}$ element is $1$. 
The transformed demand matrix $\mathbb{D}^{\prime}$ is defined as follows,
\begin{align}
\mathbb{D}^{\prime}=[x_{1,1},\ldots,x_{1,|\Lc|}; \ldots; x_{\Ksf,1},\ldots,x_{\Ksf,|\Lc|}].\label{eq:transformed demand matrix}
\end{align}

In addition, for each $i\in[|\Lc|]$ and each $\Wc \subseteq [\Ksf]$ where $|\Wc|=t$, we define
\begin{align}
 F^{\prime}_{i,\Wc}:=y_{\Lc(i),1} F_{1,\Wc} +\ldots+ y_{\Lc(i),\Nsf}  F_{\Nsf,\Wc},\label{eq:new subfiles}
\end{align}
refer  $ F^{\prime}_{i,\Wc}$ to as a {\it transformed subfile}, and refer 
$$
 B^{\prime}_{k,\Wc}=x_{k,1} F^{\prime}_{1,\Wc} +\ldots+ x_{k,|\Lc|}  F^{\prime}_{|\Lc|,\Wc}
$$
 to as a {\it transformed block}.

{\it Encoding.}
For each $\Sc\subseteq [\Ksf]$, we denote  the set of  leaders in $\Sc $   by 
\begin{align}
\Lc_{\Sc}:=\Sc \cap \Lc,\label{eq:leaders in S}
\end{align}
and the set of non-leaders in $\Sc$ by
 \begin{align}
\Nc_{\Sc}:= \Sc \setminus \Lc.\label{eq:non leaders in S}
\end{align}
We also denote  the leader indices of leaders in $\Sc $   by 
\begin{align}
\text{Ind}_{\Sc}:= \{i\in [|\Lc|]: \Lc(i)\in \Sc \},\label{eq:leaders indices in S}
\end{align} 
For example, if $\Lc=\{2,4,5\}$ and $\Sc=\{1,2,5\}$, we have $\Lc_{\Sc}=\{2,5\}$, $\Nc_{\Sc}=\{1\}$, and  $\text{Ind}_{\Sc}=\{1,3\}$.

Now we focus on  each set $\Sc \subseteq [\Ksf]$ where $|\Sc|=t+1$, and generate the multicast message
\begin{align}
  W_{\Sc}=\sum_{i\in [|\Lc_{\Sc}|]} (-1)^{i-1} B^{\prime}_{\Lc_{\Sc}(i), \Sc\setminus \left\{\Lc_{\Sc}(i)\right\}}  +\sum_{j\in [|\Nc_{\Sc}|]}(-1)^{j-1} B^{\prime}_{\Nc_{\Sc}(j),\Sc\setminus \{\Nc_{\Sc}(j)\}}. \label{eq:general multicast message in Fq}  
\end{align}
The construction of $W_{\Sc}$ can be explained as follows.
\begin{itemize}
\item The coefficient of each transformed block in $W_{\Sc}$ is either $1$ or $-1$.
\item We divide the transformed blocks in $W_{\Sc}$ into two groups, demanded by leaders and non-leaders, respectively. We alternate the sign (i.e., the coefficient $ 1$ or $-1$) of each transformed block demanded by leaders, and  alternate the sign  of each transformed block demanded by non-leaders, respectively. We then sum the resulting summations of these two groups.
\item For each $i\in [|\Lc_{\Sc}|$, by the construction in~\eqref{eq:demand transformation}, we have 
$$
B^{\prime}_{\Lc_{\Sc}(i), \Sc\setminus \left\{\Lc_{\Sc}(i)\right\}} = F^{\prime}_{\text{Ind}_{\Sc}(i), \Sc\setminus \left\{\Lc_{\Sc}(i)\right\}}.
$$
\end{itemize}

 {\it Delivery.}
The server broadcasts $W_{\Sc}$ for each $\Sc \subseteq [\Ksf]$ where $|\Sc|=t+1$ and $\Sc \cap \Lc \neq \emptyset$.

{\it Decoding.}
We consider each set $\Ac \subseteq [\Ksf]$ where $|\Ac|=t+1$ and $\Ac \cap \Lc = \emptyset$.

  We  define that   the {\it non-leader index}  of non-leader $\Ac(i)$ is $i$, where $i\in [t+1]$.
  For each $\Sc \subseteq \Ac\cup \Lc$,     
         recall that $\text{Ind}_{\Sc}$ defined in~\eqref{eq:leaders indices in S} represents the leader indices of leaders in $\Sc $  and that $\Nc_{\Sc}$ defined in~\eqref{eq:non leaders in S}
represents the set of non-leaders in $\Sc$. By   definition, we have $\Nc_{\Sc} \subseteq \Ac$. In addition, with a slight abuse of notation we denote the non-leader indices of non-leaders in $\Ac \setminus \Sc$ by 
\begin{align}
\overline{\text{Ind}}_{\Sc}=\{i\in [t+1]: \Ac(i) \notin \Sc \}.\label{eq:non leaders indices in S}
\end{align}
For example, if $\Ac=\{4,5,6\}$ and $\Sc=\{1,2,5\}$, we have  $\overline{\text{Ind}}_{\Sc}=\{1,3\}$.    

 For any set $\Xc$ and any number $y$, we define $\text{Tot}(\Xc)$ as the sum of the elements in $\Xc$, i.e., 
\begin{align}
&\text{Tot}(\Xc):=\sum_{i\in |\Xc|}\Xc(i);\label{eq:def of TOT}.
\end{align}
 For example, if $\Xc=\{1,3,4,5\}$, we have $\text{Tot}(\Xc)=1+3+4+5=13$.
 
Recall that $\mathbb{A}_{\Sc,\Vc}$ represents the sub-matrix of $\mathbb{A}$ by selecting from $\mathbb{A}$, the rows with indices in $\Sc$  and the columns with indices in $\Vc$.
It will be proved in Appendix~\ref{sec:proof of main interference cancellation Fq} that 
\begin{align}
&W_{\Ac}= \sum_{\Sc \subseteq \Ac\cup \Lc:|\Sc|=t+1, \Sc\neq \Ac} \beta_{\Ac,\Sc} W_{\Sc}, \label{eq:main interference cancellation Fq}\\
& \beta_{\Ac,\Sc}= (-1)^{1 +\text{Tot}(\overline{\text{Ind}}_{\Sc})} \text{det}(\mathbb{D}^{\prime}_{\Ac \setminus \Sc,\text{Ind}_{\Sc} }).
\label{eq:main beta}
\end{align}
 In other words, each user $k\in [\Ksf]$ can recover all messages $W_{\Sc}$ where $\Sc\subseteq [\Ksf]$ and $|\Sc|=t+1$. 
 For each desired transformed block $B^{\prime}_{k,\Wc}$, where $\Wc\subseteq ([\Ksf] \setminus\{k\})$ and $|\Wc|=t$,
 user $k$ can recover it in $W_{\Wc\cup \{k\}}$, because it knows all the other transformed blocks in $W_{\Wc\cup \{k\}}$.
 Hence, user $k$ can recover $ x_{k,1} F^{\prime}_1 +\ldots+ x_{k,|\Lc|}  F^{\prime}_{|\Lc|}$, which is identical to its demand.
 
 {\it Performance.}
In total, we transmit  $\binom{\Ksf}{t+1}-\binom{\Ksf-\text{rank}_{\qsf} (\mathbb{D}) }{t+1}$ multicast messages, each of which contains $\frac{\Bsf}{ \binom{\Ksf}{t}}$ symbols. Hence, the transmitted load is
\begin{align}
\frac{\binom{\Ksf}{t+1}-\binom{\Ksf-\text{rank}_{\qsf} (\mathbb{D}) }{t+1}}{ \binom{\Ksf}{t}}.\label{eq:load Fq for each demand}
\end{align}
 
 For the worst-case demands where $\text{rank}_{\qsf} (\mathbb{D})$ is full-rank, we have $\text{rank}_{\qsf} (\mathbb{D}) =\min\{\Ksf,\Nsf\}$, and we  achieve the worst-case load in~\eqref{eq:optimal worst case load}.

\section{Conclusions}
\label{sec:conclusion} 
In this paper, we introduced a novel problem, cache-aided function retrieval, which is a generalization of the classic coded caching problem and allows users to request scalar linear functions of files.
We proposed a novel scheme for the demands functions on arbitrary finite field. The proposed scheme was proved to be optimal under the constraint of  uncoded cache placement. In addition, for any demand, the achieved load only depends on the rank of the demand matrix.
From the results in this paper, we showed that compared to the original MAN caching problem, the optimal worst-case load of coded caching under the constraint of uncoded cache placement,  is not increased when users request scalar linear functions.

Further works include the extension of the proposed caching scheme to the case where the demanded functions are non-linear or vectorial, and finding novel caching schemes for the cache-aided function retrieval problem with  coded cache placement.


 \appendices
 
\section{Proof of Lemma~\ref{lem:F2 decodability}}
\label{proof lem F2 decodability}
To prove Lemma~\ref{lem:F2 decodability}, it is equivalent to prove that the number of  users $k\in \Bc\setminus \Wc$ satisfying the following constraints is even, 
\begin{enumerate}
\item Constraint 1: $\text{rank}_2(\mathbb{D}_{ \Bc\setminus (\Wc\cup\{k\}) })= |\Lc|$;
\item Constraint 2:  $y_{k,i} \neq 0$.
\end{enumerate}
We assume that user $k_1$ satisfies the above constraints. Hence,   $\mathbb{D}_{ \Bc\setminus (\Wc\cup\{k_1\}) }$ is full-rank, and   $y_{k_1,i} \neq 0$.
We let $\Yc=\{\Yc(1),\ldots, \Yc(|\Lc|)\}=\Bc\setminus (\Wc\cup\{k_1\}) $. 

In the following, we operate a  linear space transformation.  More precisely, we let 
\begin{align}
G_j= \yv_{\Yc(j)}[F_1;\ldots;F_{\Nsf}], \ \forall j\in [|\Lc|].\label{eq:F2 demand transformation}
\end{align}
From~\eqref{eq:F2 demand transformation}, we can re-write the demand of each user $\Yc(j)$ as 
$$
G_j=\yv^{\prime}_{j} [G_1;\ldots;G_{|\Lc|}],
$$
 where $\yv^{\prime}_{j}$ is the $|\Lc|$-dimension unit vector whose $j^{\text{th}}$ element is $1$. The transformed demand matrix of the users in $\Yc$ is 
 $$
 \mathbb{D}^{\prime}_{\Yc}=[\yv^{\prime}_{1};\ldots;\yv^{\prime}_{|\Lc|}],
 $$
 which is an identity matrix.
 
 In addition, we can also re-write the demand of user $k_1$ as 
 $$
 \yv^{\prime} [G_1;\ldots;G_{|\Lc|}],
 $$
 where $\yv^{\prime}$ is an $|\Lc|$-dimension  vector on $\mathbb{F}_2$. 
Notice that  if the $p^{\text{th}}$ element in   $\yv^{\prime}$ is $1$ and $G_p$ contains $F_{i}$, $F_{i}$ appears one time in $ \yv^{\prime} [G_1;\ldots;G_{|\Lc|}]$. 
Since $y_{k_1,i} \neq 0$,  it  can be seen that  $ \yv^{\prime} [G_1;\ldots;G_{|\Lc|}]$  contains  $F_{i}$. Hence,   the number of $p \in [|\Lc|]$  where the $p^{\text{th}}$ element in      $\yv^{\prime}$ is $1$ and $G_p$ contains $F_{i}$, is odd.
For each of such $p$, if we replace the $p^{\text{th}}$ row of $\mathbb{D}^{\prime}_{\Yc}$ by  $\yv^{\prime}$, the resulting matrix is still full-rank, because the  $p^{\text{th}}$ element in   $\yv^{\prime}$ is $1$.
Since the resulting matrix is full-rank, it can be seen that $ \mathbb{D}_{ \Bc\setminus (\Wc\cup\{\Yc(p)\}) }  $ is also full-rank.
In addition,  since $G_p$ contains $F_{i}$, we can see that $y_{\Yc(p),i} \neq 0$. Hence, user $\Yc(p)$ also satisfies the two constraints.
Moreover,  for any $s \in [|\Lc|]$, if the $s^{\text{th}}$ element  in      $\yv^{\prime}$  is not  $1$, user $\Yc(s)$  does not satisfy Constraint 1; if $G_{s}$  does not contain $F_{s}$,  user $\Yc(s)$  does not satisfy Constraint 2.

As a result, besides user $k_1$, the number of users in $\Bc\setminus \Wc$ satisfying the two constraints is odd. In conclusion, by taking user $k_1$ into consideration, the number of users in $\Bc\setminus \Wc$ satisfying the two constraints is even. Thus Lemma~\ref{lem:F2 decodability} is proved.

  \section{Proof of~\eqref{eq:main interference cancellation Fq}}
  \label{sec:proof of main interference cancellation Fq}
  We focus on one  set of non-leaders $\Ac \subseteq [\Ksf]$ where $|\Ac|=t+1$ and $\Ac \cap \Lc = \emptyset$.

For any positive integer $n$, $\text{Perm}(n)$ represents the set of all permutations of $[n]$. For any set $\Xc$ and any number $y$, we define 
$\text{Card}(\Xc,y)$ as the number of elements in $\Xc$ which is  smaller than $y$, i.e., 
\begin{align}
\text{Card}(\Xc,y):= |\{i\in \Xc:i<y\}|. \label{eq:def of NUM}
\end{align}
 For example, if $\Xc=\{1,3,4,5\}$ and $y=4$, we have    $\text{Card}(\Xc,y)=|\{1,3\}|=2$.

Our objective is to prove 
\begin{align}
&W_{\Ac}= \sum_{\Sc \subseteq \Ac\cup \Lc:|\Sc|=t+1, \Sc\neq \Ac} \beta_{\Ac,\Sc} W_{\Sc}  , \label{eq:main interference cancellation Fq app}\\
& \beta_{\Ac,\Sc}= \sum_{\substack{\uv=(u_1,\ldots,u_{|\text{Ind}_{\Sc}|})\\ \in \text{Perm}(|\text{Ind}_{\Sc}|)}} (-1)^{1 +\text{Tot}(\overline{\text{Ind}}_{\Sc})+\underset{i_1\in [|\text{Ind}_{\Sc}|]}{\sum}  \text{Card}([|\text{Ind}_{\Sc}|]\setminus\{u_1,\ldots,u_{i_1} \},u_{i_1})  } \prod_{i_2\in [|\text{Ind}_{\Sc}|]} x_{ \Ac\big(\overline{\text{Ind}}_{\Sc}(u_{i_2})\big) ,\text{Ind}_{\Sc}(i_2)},\label{eq:def of beta}
\end{align}
 where~\eqref{eq:def of beta} is obtained from expand the determinant in~\eqref{eq:main beta}.
Let us go back to the illustrated example in Section~\ref{sub:example Fq}, where we choose $\Lc= [3]$.  When $\Ac=\{4,5\}$ and $\Sc=\{1,2\}$, from the definition in~\eqref{eq:leaders indices in S} we have $\text{Ind}_{\Sc}=[2]$ and from the definition in~\eqref{eq:non leaders indices in S} we have $\overline{\text{Ind}}_{\Sc}=[2]$. In addition,  $\text{Perm}(|\text{Ind}_{\Sc}|)=\text{Perm}(2)\{(1,2),(2,1)\}$.
Hence, when $\uv =(u_1,u_2)=(1,2)$, in~\eqref{eq:def of beta} we have the term 
\begin{align}
&(-1)^{1+\text{Tot}([2])+\text{Card}([2]\setminus \{1\}, 1) + \text{Card}([2]\setminus \{1,2\}, 2)} x_{4,1} x_{5,2} \nonumber\\
&=(-1)^{1+\text{Tot}([2])+\text{Card}([2]\setminus \{1\}, 1) + \text{Card}([2]\setminus \{1,2\}, 2)} y_{4,1} y_{5,2} \nonumber\\
&= 
 y_{4,1} y_{5,2},\label{eq:first term}
\end{align}
 where~\eqref{eq:first term} is because in the example we have $F_{i}=F^{\prime}_i$ for each $i\in[\Nsf]$, and  thus $\xv_k=\yv_k$ for each $k \in [\Ksf]$.
Similarly, when $\uv=(2,1)$, in~\eqref{eq:def of beta} we have the term 
\begin{align}
&(-1)^{1+\text{Tot}([2])+\text{Card}([2]\setminus \{2\}, 2) + \text{Card}([2]\setminus \{1,2\}, 1)} x_{5,1} x_{4,2} \nonumber\\
&=(-1)^{1+\text{Tot}([2])+\text{Card}([2]\setminus \{2\}, 2) + \text{Card}([2]\setminus \{1,2\}, 1)} y_{5,1} y_{4,2} \nonumber\\
&= 
 -y_{5,1} y_{4,2}.\label{eq:second term}
\end{align}
Hence, in~\eqref{eq:def of beta} we have $\beta_{\{4,5\},\{1,2\}}= y_{4,1} y_{5,2}-y_{5,1} y_{4,2}$, which coincides~\eqref{eq:B12} in the illustrated example.

By the definition of $W_{\Sc}$ in~\eqref{eq:general multicast message in Fq},
 it is obvious to check that in~\eqref{eq:main interference cancellation Fq app}, there only exist the transformed subfiles $F_{i,\Wc}$ where $i\in [\Nsf]$, $|\Wc|\subseteq (\Ac\cup \Lc)$, and $|\Wc|=t$. 
 Now we divide such transformed subiles into hierarchies, where we say a  transformed subfile  $F^{\prime}_{i,\Wc}$ appearing in~\eqref{eq:main interference cancellation Fq app} is in Hierarchy $h\in [0:t]$, if $|\Wc\cap \Lc|=h$. In addition, on the LHS of~\eqref{eq:main interference cancellation Fq app}, only transformed subfiles in Hierarchy $0$ exist.

We consider the following three cases, 
\begin{enumerate}
\item Case 1: $F^{\prime}_{i,\Wc}$ is in Hierarchy $0$. In Appendix~\ref{sub:case 1}, we will prove that the coefficient of $F^{\prime}_{i,\Wc}$ on the RHS of~\eqref{eq:main interference cancellation Fq app} is equal to   the coefficient of $F^{\prime}_{i,\Wc}$ on the LHS of~\eqref{eq:main interference cancellation Fq app}.
\item Case 2: $F^{\prime}_{i,\Wc}$ is in Hierarchy $h>0$ and $\Lc(i)\in \Wc$. In Appendix~\ref{sub:case 2}, we will prove that the coefficient of  $F^{\prime}_{i,\Wc}$ on the RHS of~\eqref{eq:main interference cancellation Fq app} is $0$.
\item Case 3: $F^{\prime}_{i,\Wc}$ is in Hierarchy $h>0$ and $\Lc(i)\notin \Wc$.   In Appendix~\ref{sub:case 3}, we will prove that the coefficient of  $F^{\prime}_{i,\Wc}$ on the RHS of~\eqref{eq:main interference cancellation Fq app} is $0$.
\end{enumerate}
Hence, after proving the above three cases,~\eqref{eq:main interference cancellation Fq app} can be directly derived.

In the illustrated example in Section~\ref{sub:example Fq}, since  $F_{i}=F^{\prime}_i$ for each $i\in[\Nsf]$, it can be seen that $F^{\prime}_{i,\Wc}=F_{i,\Wc}$ for each  $i\in [\Nsf]$, $|\Wc|\subseteq  [\Ksf]$, and $|\Wc|=t$. For each subfile $F_{i,\Wc}$, it is in one of the following three cases, 
\begin{enumerate}
\item Case 1: $F_{i,\Wc}$ is in Hierarchy $0$.  In this case, we have the subfiles $F_{i,\{4\}}$, $F_{i,\{5\}}$ for $i\in [3]$.
\item Case 2: $F_{i,\Wc}$ is in Hierarchy $1$ and $\Lc(i)\in \Wc$. In this case, we have the subfiles $F_{i,\{i\}}$ for $i\in [3]$.
\item Case 3: $F_{i,\Wc}$ is in Hierarchy $1$ and $\Lc(i)\notin \Wc$.   In this case, we have the subfiles $F_{i,\{j\}}$ for $i\in [3]$ and $j \in [3]\setminus \{i\}$.
\end{enumerate}

\subsection{Case 1}
 \label{sub:case 1}
 If  $F^{\prime}_{i,\Wc}$ is in Hierarchy $0$, we have $\Wc \subseteq \Ac$. Since $|\Ac|-|\Wc|=1$, we assume that $\{\Ac(k)\} = \Ac \setminus \Wc$.
 On the LHS of~\eqref{eq:main interference cancellation Fq app},  $F^{\prime}_{i,\Wc}$ appears in $W_{\Ac}$, where from~\eqref{eq:general multicast message in Fq} we have 
 \begin{align}
 W_{\Ac}=\sum_{j\in [t+1]}(-1)^{j-1} (x_{\Ac(j),1} F^{\prime}_{1,\Sc\setminus \{\Ac(j)\}} +\ldots+ x_{\Ac(j),|\Lc|}  F^{\prime}_{|\Lc|,\Sc\setminus \{\Ac(j)\}} ).\label{eq:W_A in Fq}
 \end{align}
 Hence, the coefficient of  $F^{\prime}_{i,\Wc}$ in $W_{\Ac}$ is $(-1)^{k-1}x_{\Ac(k), i}$.
 
 Let us then focus on the RHS of~\eqref{eq:main interference cancellation Fq app}.         $F^{\prime}_{i,\Wc}$ appears in $W_{\Wc\cup \{\Lc(i)\} }$. Since $\Lc(i)$ is the only leader in $\Wc\cup \{\Lc(i)\}$ (i.e., $\text{Ind}_{\Wc\cup \{\Lc(i)\} }=\{i\} $),  the coefficient of  $F^{\prime}_{i,\Wc}$ in $W_{\Wc\cup \{\Lc(i)\} }$  is $(-1)^{1-1}=1$. In addition, by computing $\overline{\text{Ind}}_{\Wc\cup \{\Lc(i)\} }=\{k\}$, we have 
 \begin{align}
 \beta_{\Ac,\Wc\cup \{\Lc(i)\} }=(-1)^{1+k+0}x_{\Ac(k),i}=(-1)^{k+1} x_{\Ac(k),i}=(-1)^{k-1} x_{\Ac(k),i}.
 \end{align}
 Hence, the coefficient of  $F^{\prime}_{i,\Wc}$ on the RHS of~\eqref{eq:main interference cancellation Fq app} (i.e., in $ \beta_{\Ac,\Wc\cup \{\Lc(i)\} } W_{\Wc\cup \{\Lc(i)\} }$) is 
\begin{align}
(-1)^{k-1} x_{\Ac(k),i} \times 1=(-1)^{k-1} x_{\Ac(k),i},
\end{align} 
 which is the same as the coefficient of  $F^{\prime}_{i,\Wc}$ on the LHS of~\eqref{eq:main interference cancellation Fq app}. 
 
 \subsection{Case 2}
 \label{sub:case 2}
 Now we focus on one transformed subfile $F^{\prime}_{i,\Wc}$  in Hierarchy $h>0$ where  $\Lc(i) \in \Wc$. By definition, we have $|\Wc \cap \Lc|=h$. On the RHS of~\eqref{eq:main interference cancellation Fq app}, since $\Lc(i) \in \Wc$, $F^{\prime}_{i,\Wc}$ only appears in $W_{\Wc\cup \{\Ac(k)\}}$, where  $k\in \overline{\text{Ind}}_{\Wc}$. 
We define that 
\begin{align}
\text{the $\left( \overline{\text{Ind}}^{-1}_{\Wc}(k)\right)^{\text{th}}$ smallest element in  $\overline{\text{Ind}}_{\Wc}$ is $k$.} \label{eq:def of ind bar -1}
\end{align}

We focus on one $k\in \overline{\text{Ind}}_{\Wc}$.  $\Ac(k)$ is the $k^{\text{th}}$ element in $\Ac$, and  in $\Ac \setminus \Wc$ there are $ \overline{\text{Ind}}^{-1}_{\Wc}(k)-1$ elements smaller than $\Ac(k)$. Hence, 
in $\Nc_{\Wc \cup \{\Ac(k)\}}$ there are $k-1- \left( \overline{\text{Ind}}^{-1}_{\Wc}(k)-1 \right)=k- \overline{\text{Ind}}^{-1}_{\Wc}(k)$ elements smaller than $\Ac(k)$. So 
from~\eqref{eq:general multicast message in Fq}, it can be seen that the coefficient of  $F^{\prime}_{i,\Wc}$ in $W_{\Wc\cup \{\Ac(k)\}}$  is 
\begin{align}
(-1)^{k- \overline{\text{Ind}}^{-1}_{\Wc}(k)} x_{\Ac(k),i}.\label{eq:encoding cof case 2}
\end{align}
In addition,  we have 
\begin{subequations}
 \begin{align}
& \beta_{\Ac,\Wc \cup \{\Ac(k)\} }= \sum_{\substack{\uv=(u_1,\ldots,u_{|\text{Ind}_{\Wc \cup \{\Ac(k)\}}|})\\ \in \text{Perm}(|\text{Ind}_{\Wc \cup \{\Ac(k)\}}|)}} (-1)^{1 +\text{Tot}(\overline{\text{Ind}}_{\Wc \cup \{\Ac(k)\}})+ \negmedspace\negmedspace\negmedspace \underset{i_1\in [|\text{Ind}_{\Wc\cup \{\Ac(k)\}}|]}{\sum}\negmedspace\negmedspace\negmedspace  \text{Card}([|\text{Ind}_{\Wc \cup \{\Ac(k)\}}|]\setminus\{u_1,\ldots,u_{i_1} \},u_{i_1})  } \nonumber\\& \prod_{i_2\in [|\text{Ind}_{\Wc \cup \{\Ac(k)\}}|]} x_{ \Ac\big(\overline{\text{Ind}}_{\Wc \cup \{\Ac(k)\}}(u_{i_2})\big) ,\text{Ind}_{\Wc \cup \{\Ac(k)\}}(i_2)}\\
& =\sum_{\substack{\uv=(u_1,\ldots,u_{|\text{Ind}_{\Wc}|})\\ \in \text{Perm}(|\text{Ind}_{\Wc}|)}} (-1)^{1 +\left(\text{Tot}(\overline{\text{Ind}}_{\Wc})-k\right)+\negmedspace\negmedspace\negmedspace \underset{i_1\in [|\text{Ind}_{\Wc}|]}{\sum} \negmedspace\negmedspace\negmedspace \text{Card}([|\text{Ind}_{\Wc}|]\setminus\{u_1,\ldots,u_{i_1} \},u_{i_1})  } \prod_{i_2\in [|\text{Ind}_{\Wc}|]} x_{ \Ac\big(\overline{\text{Ind}}_{\Wc \cup \{\Ac(k)\}}(u_{i_2})\big) ,\text{Ind}_{\Wc}(i_2)}.\label{eq:decoding cof case 2}
 \end{align}
\end{subequations}
From~\eqref{eq:encoding cof case 2} and~\eqref{eq:decoding cof case 2}, the coefficient of $F^{\prime}_{i,\Wc}$ in $\beta_{\Ac,\Wc \cup \{\Ac(k)\} }  W_{\Wc\cup \{\Ac(k)\}}$ is  $(-1)^{k- \overline{\text{Ind}}^{-1}_{\Wc}(k)} x_{\Ac(k),i} \beta_{\Ac,\Wc \cup \{\Ac(k)\} }$.

In the following, we will prove 
\begin{align}
\sum_{k\in \overline{\text{Ind}}_{\Wc}} (-1)^{k- \overline{\text{Ind}}^{-1}_{\Wc}(k)} x_{\Ac(k),i} \beta_{\Ac,\Wc \cup \{\Ac(k)\} }=0,\label{eq:to prove case 2}
\end{align}
such that the coefficient of $F^{\prime}_{i,\Wc}$  on the RHS of~\eqref{eq:main interference cancellation Fq app} is $0$.

Let us focus on one $k \in \overline{\text{Ind}}_{\Wc}$ and one permutation $\uv=(u_1,\ldots,u_{|\text{Ind}_{\Wc}|}) \in \text{Perm}(|\text{Ind}_{\Wc}|) $. The term in~\eqref{eq:to prove case 2} caused by $k $ and $\uv$ is 
\begin{subequations}
\begin{align}
&(-1)^{k - \overline{\text{Ind}}^{-1}_{\Wc}(k )} x_{\Ac(k ),i}  \negmedspace \left\{  \negmedspace  (-1)^{1 +\left(\text{Tot}(\overline{\text{Ind}}_{\Wc})-k\right)+\negmedspace\negmedspace\negmedspace\underset{i_1\in [|\text{Ind}_{\Wc}|]}{\sum} \negmedspace\negmedspace\negmedspace \text{Card}([|\text{Ind}_{\Wc}|]\setminus\{u_1,\ldots,u_{i_1} \},u_{i_1})  }\negmedspace \negmedspace\negmedspace\negmedspace \prod_{i_2\in [|\text{Ind}_{\Wc}|]}  \negmedspace\negmedspace\negmedspace\negmedspace x_{ \Ac\big(\overline{\text{Ind}}_{\Wc \cup \{\Ac(k)\}}(u_{i_2})\big) ,\text{Ind}_{\Wc}(i_2)}  \negmedspace \right\} \\
&=(-1)^{ - \overline{\text{Ind}}^{-1}_{\Wc}(k ) +1+\text{Tot}(\overline{\text{Ind}}_{\Wc}) +\negmedspace\negmedspace\negmedspace \underset{i_1\in [|\text{Ind}_{\Wc}|]}{\sum}\negmedspace\negmedspace\negmedspace \text{Card}([|\text{Ind}_{\Wc}|]\setminus\{u_1,\ldots,u_{i_1} \},u_{i_1})  }\negmedspace \left\{ \negmedspace x_{\Ac(k ),i} \negmedspace\prod_{i_2\in [|\text{Ind}_{\Wc}|]}\negmedspace x_{ \Ac\big(\overline{\text{Ind}}_{\Wc \cup \{\Ac(k)\}}(u_{i_2})\big) ,\text{Ind}_{\Wc}(i_2)}\right\}. \label{eq:case 2 term k u}
\end{align}
\end{subequations}
Notice that in the product  
\begin{align}
x_{\Ac(k ),i} \prod_{i_2\in [|\text{Ind}_{\Wc}|]}  x_{ \Ac\big(\overline{\text{Ind}}_{\Wc \cup \{\Ac(k)\}}(u_{i_2})\big) ,\text{Ind}_{\Wc}(i_2)}, \label{eq:case 2 considered product}
\end{align}
there is one term whose second subscript is $i^{\prime}$ for each $i^{\prime} \in \text{Ind}_{\Wc}\setminus \{i\} $, and there are two terms whose second subscript is $i$.   We define that 
\begin{align}
\text{the $\left( \text{Ind}_{\Wc}^{-1}(i)\right)^{\text{th}}$ smallest element in  $\text{Ind}_{\Wc}$ is $i$.}\label{eq:def of ind -1}
\end{align}
 Hence, the two terms in~\eqref{eq:case 2 considered product} whose second subscript is $i$ are $x_{\Ac(k ),i}$ and $x_{ \Ac(k^{\prime}) ,i}$, where
 $k^{\prime}:= \overline{\text{Ind}}_{\Wc \cup \{\Ac(k)\}}(u_{ \text{Ind}_{\Wc}^{-1}(i) })  $.
 
In addition,  the combination $k^{\prime}$ and $\uv^{\prime}=(u^{\prime}_1,\ldots,u^{\prime}_{|\text{Ind}_{\Wc}|})$ also causes a term in~\eqref{eq:to prove case 2} 
which has the product 
\begin{align}
x_{\Ac(k^{\prime}),i} \prod_{i_2\in [|\text{Ind}_{\Wc}|]}  x_{ \Ac\big(\overline{\text{Ind}}_{\Wc \cup \{\Ac(k^{\prime})\}}(u^{\prime}_{i_2})\big) ,\text{Ind}_{\Wc}(i_2)}. \label{eq:case 2 considered product kprime}
\end{align}
The products in~\eqref{eq:case 2 considered product} and~\eqref{eq:case 2 considered product kprime} are identical if $\uv^{\prime}$ is as follows, 
\begin{itemize}
\item for $j \in [|\text{Ind}_{\Wc}|]\setminus \{ \text{Ind}_{\Wc}^{-1}(i) \}$, we have 
\begin{align}
\Ac\big(\overline{\text{Ind}}_{\Wc \cup \{\Ac(k^{\prime})\}}(u^{\prime}_j)\big)= \Ac\big(\overline{\text{Ind}}_{\Wc \cup \{\Ac(k)\}}(u_{j})\big) ;\label{eq:case 2 other terms}
\end{align}
such that 
\begin{align}
x_{ \Ac\big(\overline{\text{Ind}}_{\Wc \cup \{\Ac(k^{\prime})\}}(u^{\prime}_{j})\big) ,\text{Ind}_{\Wc}(j)}= x_{ \Ac\big(\overline{\text{Ind}}_{\Wc \cup \{\Ac(k)\}}(u_{j})\big) ,\text{Ind}_{\Wc}(j)};
\end{align} 
 \item for $j=\text{Ind}_{\Wc}^{-1}(i)$, we have 
 \begin{align}
 \Ac\big(\overline{\text{Ind}}_{\Wc \cup \{\Ac(k^{\prime})\}}(u^{\prime}_j)\big)=\Ac(k);\label{eq:case 2 mid term}
 \end{align}
 such that 
 \begin{align}
 x_{ \Ac\big(\overline{\text{Ind}}_{\Wc \cup \{\Ac(k^{\prime})\}}(u^{\prime}_{j})\big) ,\text{Ind}_{\Wc}(j)}=x_{\Ac(k),i}.
 \end{align}
\end{itemize}

 
 It is obvious to check that  there does not exist any other combination of $k^{\prime\prime} \in  \overline{\text{Ind}}_{\Wc}$ and $\uv^{\prime\prime} \in \text{Perm}(|\text{Ind}_{\Wc}|)$,   causing a term on the LHS of~\eqref{eq:to prove case 2} which has the product  in~\eqref{eq:case 2 considered product}, except the two above combinations.

In Appendix~\ref{sec:proof of eq:case 2 cancel two term}, we will prove that 
\begin{align}
& (-1)^{ - \overline{\text{Ind}}^{-1}_{\Wc}(k ) +1+\text{Tot}(\overline{\text{Ind}}_{\Wc}) +\negmedspace\negmedspace\negmedspace \underset{i_1\in [|\text{Ind}_{\Wc}|]}{\sum}\negmedspace\negmedspace\negmedspace \text{Card}([|\text{Ind}_{\Wc}|]\setminus\{u_1,\ldots,u_{i_1} \},u_{i_1})  }  + \nonumber\\&   (-1)^{ - \overline{\text{Ind}}^{-1}_{\Wc}(k^{\prime} ) +1+\text{Tot}(\overline{\text{Ind}}_{\Wc}) +\negmedspace\negmedspace\negmedspace \underset{i^{\prime}_1\in [|\text{Ind}_{\Wc}|]}{\sum}\negmedspace\negmedspace\negmedspace \text{Card}([|\text{Ind}_{\Wc}|]\setminus\{u^{\prime}_1,\ldots,u^{\prime}_{i^{\prime}_1} \},u^{\prime}_{i^{\prime}_1})  } =0,\label{eq:case 2 cancel two term}
\end{align}  
   such that the coefficient of the product in~\eqref{eq:case 2 considered product} on the LHS of~\eqref{eq:to prove case 2} is $0$. In other words,  for each combination of $k$ and $\uv$  on the LHS of~\eqref{eq:to prove case 2}, there is exactly one term caused by the combination of $k^{\prime}$ and  $\uv^{\prime}$, such that
     the sum of these two caused terms is $0$. Thus~\eqref{eq:to prove case 2} is proved.

 \subsection{Case 3}
 \label{sub:case 3}
Lastly we focus on one transformed subfile $F^{\prime}_{i,\Wc}$  in Hierarchy $h>0$ where  $\Lc(i) \notin \Wc$. By definition, we have $|\Wc \cap \Lc|=h$. On the RHS of~\eqref{eq:main interference cancellation Fq app}, since $\Lc(i) \notin \Wc$,
 $F^{\prime}_{i,\Wc}$ appears in $W_{\Wc\cup \{\Lc(i)\}}$. In addition,  $F^{\prime}_{i,\Wc}$ also appears 
in $W_{\Wc\cup \{\Ac(k)\}}$, where  $k\in \overline{\text{Ind}}_{\Wc}$. 

 Let us first focus on $W_{\Wc\cup \{\Lc(i)\}}$. Recall that the $\left( \text{Ind}_{\Wc \cup\{\Lc(i)\} }^{-1}(i)\right)^{\text{th}}$ element in  $\text{Ind}_{\Wc \cup\{\Lc(i)\}}$ is $i$.
From~\eqref{eq:general multicast message in Fq}, it can be seen that the coefficient of  $F^{\prime}_{i,\Wc}$ in $W_{\Wc\cup \{\Lc(i)\}}$  is 
\begin{align}
(-1)^{ \text{Ind}_{\Wc \cup\{\Lc(i)\} }^{-1}(i)-1 } .\label{eq:encoding cof leader case 3}
\end{align}
 In addition, we have
 \begin{subequations}
 \begin{align}
 & \beta_{\Ac,\Wc \cup \{\Lc(i)\} }  =\sum_{\substack{\uv=(u_1,\ldots,u_{|\text{Ind}_{\Wc \cup \{\Lc(i)\}}|})\\ \in \text{Perm}(|\text{Ind}_{\Wc \cup \{\Lc(i)\}}|)}} (-1)^{1+\text{Tot}(\overline{\text{Ind}}_{\Wc \cup \{\Lc(i)\}})+ \negmedspace\negmedspace\negmedspace \underset{i_1\in [|\text{Ind}_{\Wc\cup \{\Lc(i)\}}|]}{\sum}\negmedspace\negmedspace\negmedspace  \text{Card}([|\text{Ind}_{\Wc \cup \{\Lc(i)\}}|]\setminus\{u_1,\ldots,u_{i_1} \},u_{i_1})  } \nonumber\\& \prod_{i_2\in [|\text{Ind}_{\Wc \cup \{\Lc(i)\}}|]} x_{ \Ac\big(\overline{\text{Ind}}_{\Wc \cup \{\Lc(i)\}}(u_{i_2})\big) ,\text{Ind}_{\Wc \cup \{\Lc(i)\}}(i_2)}\\
 &= \sum_{\substack{\uv=(u_1,\ldots,u_{|\text{Ind}_{\Wc }|+1})\\ \in \text{Perm}(|\text{Ind}_{\Wc }|+1)}}    \negmedspace\negmedspace\negmedspace          (-1)^{ 1 +\text{Tot}(\overline{\text{Ind}}_{\Wc})+ \negmedspace\negmedspace\negmedspace \underset{i_1\in [|\text{Ind}_{\Wc }|+1]}{\sum}\negmedspace\negmedspace\negmedspace  \text{Card}([|\text{Ind}_{\Wc }|+1]\setminus\{u_1,\ldots,u_{i_1} \},u_{i_1})  } \negmedspace\negmedspace\negmedspace  \prod_{i_2\in [|\text{Ind}_{\Wc }|+1]} \negmedspace\negmedspace\negmedspace  x_{ \Ac\big(\overline{\text{Ind}}_{\Wc  }(u_{i_2})\big) ,\text{Ind}_{\Wc \cup \{\Lc(i)\}}(i_2)}
 \end{align}
  \end{subequations}

 Let us then focus on  $W_{\Wc\cup \{\Ac(k)\}}$, where  $k\in \overline{\text{Ind}}_{\Wc}$. 
It was proved in~\eqref{eq:encoding cof case 2} that  the coefficient of  $F^{\prime}_{i,\Wc}$ in $W_{\Wc\cup \{\Ac(k)\}}$  is 
\begin{align}
(-1)^{k- \overline{\text{Ind}}^{-1}_{\Wc}(k)} x_{\Ac(k),i}.\label{eq:encoding nonleader cof case 3}
\end{align}
In addition,  it was proved in~\eqref{eq:decoding cof case 2} that 
  \begin{align}
& \beta_{\Ac,\Wc \cup \{\Ac(k)\} }  = \negmedspace\negmedspace\negmedspace \sum_{\substack{\uv=(u_1,\ldots,u_{|\text{Ind}_{\Wc}|})\\ \in \text{Perm}(|\text{Ind}_{\Wc}|)}} \negmedspace\negmedspace\negmedspace\negmedspace\negmedspace\negmedspace (-1)^{1+\left(\text{Tot}(\overline{\text{Ind}}_{\Wc})-k\right)+\negmedspace\negmedspace\negmedspace \underset{i_1\in [|\text{Ind}_{\Wc}|]}{\sum} \negmedspace\negmedspace\negmedspace \text{Card}([|\text{Ind}_{\Wc}|]\setminus\{u_1,\ldots,u_{i_1} \},u_{i_1})  } \negmedspace\negmedspace\negmedspace \prod_{i_2\in [|\text{Ind}_{\Wc}|]}  \negmedspace\negmedspace\negmedspace x_{ \Ac\big(\overline{\text{Ind}}_{\Wc \cup \{\Ac(k)\}}(u_{i_2})\big) ,\text{Ind}_{\Wc}(i_2)}.\label{eq:decoding nonleader cof case 2}
 \end{align}

 In the following, we will prove 
\begin{align}
(-1)^{ \text{Ind}_{\Wc \cup\{\Lc(i)\} }^{-1}(i)-1 } \beta_{\Ac,\Wc \cup \{\Lc(i)\} } +      \sum_{k\in \overline{\text{Ind}}_{\Wc}} (-1)^{k- \overline{\text{Ind}}^{-1}_{\Wc}(k)} x_{\Ac(k),i} \beta_{\Ac,\Wc \cup \{\Ac(k)\} }=0,\label{eq:to prove case 3}
\end{align}
such that the coefficient of $F^{\prime}_{i,\Wc}$  on the RHS of~\eqref{eq:main interference cancellation Fq app} is $0$. Notice that there are $t-|\text{Ind}_{\Wc}|$ non-leaders in $\Wc$. Since there are totally $t+1$ non-leaders in $\Ac$, we have 
\begin{align}
|\overline{\text{Ind}}_{\Wc}|=t+1-(t-|\text{Ind}_{\Wc}|)=|\text{Ind}_{\Wc}|+1.\label{eq:card of bar ind W}
\end{align}

 Let us focus on one permutation $\uv=(u_1,\ldots,u_{|\text{Ind}_{\Wc }|+1}) \in \text{Perm}(|\text{Ind}_{\Wc }|+1)$ in $\beta_{\Ac,\Wc \cup \{\Lc(i)\} }$. The term in~\eqref{eq:to prove case 3} caused by $\uv$ is 
\begin{subequations}
\begin{align}
&(-1)^{ \text{Ind}_{\Wc \cup\{\Lc(i)\} }^{-1}(i)-1 }   \negmedspace \left\{   (-1)^{ 1 +\text{Tot}(\overline{\text{Ind}}_{\Wc})+ \negmedspace\negmedspace\negmedspace \underset{i_1\in [|\text{Ind}_{\Wc }|+1]}{\sum}\negmedspace\negmedspace\negmedspace  \text{Card}([|\text{Ind}_{\Wc }|+1]\setminus\{u_1,\ldots,u_{i_1} \},u_{i_1})  } \negmedspace\negmedspace\negmedspace  \prod_{i_2\in [|\text{Ind}_{\Wc }|+1]} \negmedspace\negmedspace\negmedspace  x_{ \Ac\big(\overline{\text{Ind}}_{\Wc  }(u_{i_2})\big) ,\text{Ind}_{\Wc \cup \{\Lc(i)\}}(i_2)} \right\} \\
 &=(-1)^{ \text{Ind}_{\Wc \cup\{\Lc(i)\} }^{-1}(i)   +\text{Tot}(\overline{\text{Ind}}_{\Wc})+ \negmedspace\negmedspace\negmedspace \underset{i_1\in [|\text{Ind}_{\Wc }|+1]}{\sum}\negmedspace\negmedspace\negmedspace  \text{Card}([|\text{Ind}_{\Wc }|+1]\setminus\{u_1,\ldots,u_{i_1} \},u_{i_1})   } \left\{  \prod_{i_2\in [|\text{Ind}_{\Wc }|+1]} \negmedspace\negmedspace\negmedspace  x_{ \Ac\big(\overline{\text{Ind}}_{\Wc  }(u_{i_2})\big) ,\text{Ind}_{\Wc \cup \{\Lc(i)\}}(i_2)} \right\} . \label{eq:case 3 term u}
\end{align}
\end{subequations}
We can rewrite the product term in~\eqref{eq:case 3 term u} as follows (recall again that  the $\left( \text{Ind}_{\Wc \cup\{\Lc(i)\} }^{-1}(i)\right)^{\text{th}}$ element in  $\text{Ind}_{\Wc \cup\{\Lc(i)\}}$ is $i$),
\begin{subequations}
\begin{align}
&\prod_{i_2\in [|\text{Ind}_{\Wc }|+1]} \negmedspace\negmedspace\negmedspace  x_{ \Ac\big(\overline{\text{Ind}}_{\Wc  }(u_{i_2})\big) ,\text{Ind}_{\Wc \cup \{\Lc(i)\}}(i_2)} \nonumber\\
&=x_{ \Ac\big(\overline{\text{Ind}}_{\Wc  }(u_{\text{Ind}_{\Wc \cup\{\Lc(i)\} }^{-1}(i) })\big) , i } 
\prod_{i_2\in [|\text{Ind}_{\Wc }|+1] \setminus \{\text{Ind}_{\Wc \cup\{\Lc(i)\} }^{-1}(i)\} } \negmedspace\negmedspace\negmedspace  x_{ \Ac\big(\overline{\text{Ind}}_{\Wc  }(u_{i_2})\big) ,\text{Ind}_{\Wc \cup \{\Lc(i)\}}(i_2)}\\
&=x_{ \Ac(\widetilde{k} ) , i } 
\prod_{i_2\in [|\text{Ind}_{\Wc }|+1] \setminus \{\text{Ind}_{\Wc \cup\{\Lc(i)\} }^{-1}(i)\} } \negmedspace\negmedspace\negmedspace  x_{ \Ac\big(\overline{\text{Ind}}_{\Wc  }(u_{i_2})\big) ,\text{Ind}_{\Wc \cup \{\Lc(i)\}}(i_2)},\label{eq:one product case 3}
\end{align}
 \end{subequations}
where we define $\widetilde{k} :=\overline{\text{Ind}}_{\Wc  }(u_{\text{Ind}_{\Wc \cup\{\Lc(i)\} }^{-1}(i) })$.

Hence, on the LHS of~\eqref{eq:to prove case 3},
besides  $(-1)^{ \text{Ind}_{\Wc \cup\{\Lc(i)\} }^{-1}(i)-1 } \beta_{\Ac,\Wc \cup \{\Lc(i)\} }$,
only the caused term by the combination of $\widetilde{k}$ and $\widetilde{\uv}=(\widetilde{u}_1,\ldots,\widetilde{u}_{|\text{Ind}_{\Wc}|})$ has  the product in~\eqref{eq:one product case 3}, where 
\begin{subequations}
\begin{align}
&\widetilde{\uv}=(g(u_1),\ldots,g(u_{\text{Ind}_{\Wc \cup\{\Lc(i)\} }^{-1}(i)-1}),g(u_{\text{Ind}_{\Wc \cup\{\Lc(i)\} }^{-1}(i)+1}),\ldots, g(u_{|\text{Ind}_{\Wc }|+1}) ),\label{eq:u tilde}\\
& g(u_j):=\begin{cases} u_j, & \text{ if }  u_j<  u_{\text{Ind}_{\Wc \cup\{\Lc(i)\} }^{-1}(i) } \\u_j-1 & \text{ if }  u_j>  u_{\text{Ind}_{\Wc \cup\{\Lc(i)\} }^{-1}(i) }\end{cases}.\label{eq:g function}
\end{align}
 \end{subequations}

In Appendix~\ref{sec:proof of eq:case 3 cancel two term}, we will prove that 
\begin{align}
&(-1)^{ \text{Ind}_{\Wc \cup\{\Lc(i)\} }^{-1}(i)   +\text{Tot}(\overline{\text{Ind}}_{\Wc})+ \negmedspace\negmedspace\negmedspace \underset{i_1\in [|\text{Ind}_{\Wc }|+1]}{\sum}\negmedspace\negmedspace\negmedspace  \text{Card}([|\text{Ind}_{\Wc }|+1]\setminus\{u_1,\ldots,u_{i_1} \},u_{i_1})   } + \nonumber\\
& (-1)^{ - \overline{\text{Ind}}^{-1}_{\Wc}(\widetilde{k}  ) +1+\text{Tot}(\overline{\text{Ind}}_{\Wc}) +\negmedspace\negmedspace\negmedspace \underset{\widetilde{i}_1\in [|\text{Ind}_{\Wc}|]}{\sum}\negmedspace\negmedspace\negmedspace \text{Card}([|\text{Ind}_{\Wc}|]\setminus\{\widetilde{u}_1,\ldots,\widetilde{u}_{\widetilde{i}_1} \},\widetilde{u}_{\widetilde{i}_1})  }   =0,\label{eq:case 3 cancel two term}
\end{align}  
   such that the coefficient of the product in~\eqref{eq:one product case 3} on the LHS of~\eqref{eq:to prove case 3} is $0$.
Hence, for each permutation $\uv\in \text{Perm}(|\text{Ind}_{\Wc }|+1)$, there is exactly one term caused by the combination of $\widetilde{k} \in \overline{\text{Ind}}_{\Wc} $ and $\widetilde{\uv} \in \text{Perm}(|\text{Ind}_{\Wc}|) $, such that  the sum of these two caused terms are $0$. 

   In addition, on the LHS of~\eqref{eq:to prove case 3}, there are $(|\text{Ind}_{\Wc }|+1)!$ terms in $(-1)^{ \text{Ind}_{\Wc \cup\{\Lc(i)\} }^{-1}(i)-1 } \beta_{\Ac,\Wc \cup \{\Lc(i)\} }$.   
Recall that in~\eqref{eq:card of bar ind W}, we proved    $|\overline{\text{Ind}}_{\Wc}|=|\text{Ind}_{\Wc}|+1$. Hence,  on the LHS of~\eqref{eq:to prove case 3},   there are $|\text{Ind}_{\Wc }|! \times (|\text{Ind}_{\Wc}|+1)= (|\text{Ind}_{\Wc }|+1)!$ terms in $ \sum_{k\in \overline{\text{Ind}}_{\Wc}} (-1)^{k- \overline{\text{Ind}}^{-1}_{\Wc}(k)} x_{\Ac(k),i} \beta_{\Ac,\Wc \cup \{\Ac(k)\} }$.
   In conclusion, we prove~\eqref{eq:to prove case 3}.

 \section{Proof of~\eqref{eq:case 2 cancel two term}}
 \label{sec:proof of eq:case 2 cancel two term}
 To prove~\eqref{eq:case 2 cancel two term}, it is equivalent to prove
     \begin{align}
  (-1)^{ -\overline{\text{Ind}}^{-1}_{\Wc}(k ) - \overline{\text{Ind}}^{-1}_{\Wc}(k^{\prime} )  +\negmedspace\negmedspace\negmedspace \underset{i_1\in [|\text{Ind}_{\Wc}|]}{\sum}\negmedspace\negmedspace\negmedspace \text{Card}([|\text{Ind}_{\Wc}|]\setminus\{u_1,\ldots,u_{i_1} \},u_{i_1})+\negmedspace\negmedspace\negmedspace \underset{i^{\prime}_1\in [|\text{Ind}_{\Wc}|]}{\sum}\negmedspace\negmedspace\negmedspace \text{Card}([|\text{Ind}_{\Wc}|]\setminus\{u^{\prime}_1,\ldots,u^{\prime}_{i^{\prime}_1} \},u^{\prime}_{i^{\prime}_1}) }=-1.\label{eq:case 2 equivalent to prove} 
     \end{align}

     Let us focus on $\underset{i_1\in [|\text{Ind}_{\Wc}|]}{\sum}\negmedspace\negmedspace\negmedspace \text{Card}([|\text{Ind}_{\Wc}|]\setminus\{u_1,\ldots,u_{i_1} \},u_{i_1})$. By the definition of the function $\text{Card}(\cdot)$ in~\eqref{eq:def of NUM}, we have 
     \begin{subequations}
     \begin{align}
&     \underset{i_1\in [|\text{Ind}_{\Wc}|]}{\sum}\negmedspace\negmedspace\negmedspace \text{Card}([|\text{Ind}_{\Wc}|]\setminus\{u_1,\ldots,u_{i_1} \},u_{i_1})\nonumber \\
&=\underset{i_1\in [|\text{Ind}_{\Wc}|]: i_1 \neq \text{Ind}_{\Wc}^{-1}(i) }{\sum}\negmedspace\negmedspace\negmedspace \text{Card}\big(([|\text{Ind}_{\Wc}|]\setminus \{u_{\text{Ind}_{\Wc}^{-1}(i)}\}) \setminus\{u_1,\ldots,u_{\text{Ind}_{\Wc}^{-1}(i)-1},u_{\text{Ind}_{\Wc}^{-1}(i)+1},\ldots, u_{i_1} \},u_{i_1} \big)\nonumber\\
&+|\{i_2 \in [ \text{Ind}_{\Wc}^{-1}(i) -1] : u_{\text{Ind}_{\Wc}^{-1}(i)}< u_{i_2} \}|+|\{i_3\in [\text{Ind}_{\Wc}^{-1}(i)+1:|\text{Ind}_{\Wc}| ]: u_{i_3}<u_{\text{Ind}_{\Wc}^{-1}(i)} \}|\\
&=\underset{i_1\in [|\text{Ind}_{\Wc}|]: i_1 \neq \text{Ind}_{\Wc}^{-1}(i) }{\sum}\negmedspace\negmedspace\negmedspace \text{Card}\big(([|\text{Ind}_{\Wc}|]\setminus \{u_{\text{Ind}_{\Wc}^{-1}(i)}\}) \setminus\{u_1,\ldots,u_{\text{Ind}_{\Wc}^{-1}(i)-1},u_{\text{Ind}_{\Wc}^{-1}(i)+1},\ldots, u_{i_1} \},u_{i_1} \big)\nonumber\\
&+ (\text{Ind}_{\Wc}^{-1}(i)-1- |\{i_2 \in [ \text{Ind}_{\Wc}^{-1}(i) -1] :  u_{i_2}<u_{\text{Ind}_{\Wc}^{-1}(i)} \}|) +|\{i_3\in [\text{Ind}_{\Wc}^{-1}(i)+1:|\text{Ind}_{\Wc}| ]: u_{i_3}<u_{\text{Ind}_{\Wc}^{-1}(i)} \}|\\
&=\underset{i_1\in [|\text{Ind}_{\Wc}|]: i_1 \neq \text{Ind}_{\Wc}^{-1}(i)}{\sum}\negmedspace\negmedspace\negmedspace \text{Card}\big(([|\text{Ind}_{\Wc}|]\setminus \{u_{\text{Ind}_{\Wc}^{-1}(i)}\}) \setminus\{u_1,\ldots,u_{\text{Ind}_{\Wc}^{-1}(i)-1},u_{\text{Ind}_{\Wc}^{-1}(i)+1},\ldots, u_{i_1} \},u_{i_1} \big)\nonumber\\
&+ (\text{Ind}_{\Wc}^{-1}(i)-1- |\{i_2 \in [ \text{Ind}_{\Wc}^{-1}(i) -1] :  u_{i_2}<u_{\text{Ind}_{\Wc}^{-1}(i)} \}|) \nonumber\\
&+\text{Card}([|\text{Ind}_{\Wc}|] , u_{\text{Ind}_{\Wc}^{-1}(i)})-|\{i_2 \in [ \text{Ind}_{\Wc}^{-1}(i) -1] :  u_{i_2}<u_{\text{Ind}_{\Wc}^{-1}(i)} \}|.\label{eq:case 2 num k}
     \end{align}
      \end{subequations}
    Similarly, for  $\underset{i^{\prime}_1\in [|\text{Ind}_{\Wc}|]}{\sum}\negmedspace\negmedspace\negmedspace \text{Card}([|\text{Ind}_{\Wc}|]\setminus\{u^{\prime}_1,\ldots,u^{\prime}_{i^{\prime}_1} \},u^{\prime}_{i^{\prime}_1})$, from the same derivation as~\eqref{eq:case 2 num k}, we have
    \begin{align}
    & \underset{i^{\prime}_1\in [|\text{Ind}_{\Wc}|]}{\sum}\negmedspace\negmedspace\negmedspace \text{Card}([|\text{Ind}_{\Wc}|]\setminus\{u^{\prime}_1,\ldots,u^{\prime}_{i^{\prime}_1} \},u^{\prime}_{i^{\prime}_1}) \nonumber\\
    &=\underset{i^{\prime}_1\in[|\text{Ind}_{\Wc}|]: i^{\prime}_1 \neq \text{Ind}_{\Wc}^{-1}(i) }{\sum}\negmedspace\negmedspace\negmedspace \text{Card}\big(([|\text{Ind}_{\Wc}|]\setminus \{u^{\prime}_{\text{Ind}_{\Wc}^{-1}(i)}\}) \setminus\{u^{\prime}_1,\ldots,u^{\prime}_{\text{Ind}_{\Wc}^{-1}(i)-1},u^{\prime}_{\text{Ind}_{\Wc}^{-1}(i)+1},\ldots, u^{\prime}_{i_1} \},u^{\prime}_{i_1} \big)\nonumber\\
  &+ (\text{Ind}_{\Wc}^{-1}(i)-1- |\{i^{\prime}_2 \in [ \text{Ind}_{\Wc}^{-1}(i) -1] :  u^{\prime}_{i^{\prime}_2}<u^{\prime}_{\text{Ind}_{\Wc}^{-1}(i)} \}|) \nonumber\\
&+\text{Card}([|\text{Ind}_{\Wc}|], u^{\prime}_{\text{Ind}_{\Wc}^{-1}(i)})-|\{i^{\prime}_2 \in [ \text{Ind}_{\Wc}^{-1}(i) -1] :  u^{\prime}_{i^{\prime}_2}<u^{\prime}_{\text{Ind}_{\Wc}^{-1}(i)} \}|.\label{eq:case 2 num k prime}
    \end{align}
    
In addition, from~\eqref{eq:case 2 other terms}, it can be seen that 
\begin{align}
&\underset{i_1\in [|\text{Ind}_{\Wc}|]: i_1 \neq \text{Ind}_{\Wc}^{-1}(i) }{\sum}\negmedspace\negmedspace\negmedspace \text{Card}\big(([|\text{Ind}_{\Wc}|]\setminus \{u_{\text{Ind}_{\Wc}^{-1}(i)}\}) \setminus\{u_1,\ldots,u_{\text{Ind}_{\Wc}^{-1}(i)-1},u_{\text{Ind}_{\Wc}^{-1}(i)+1},\ldots, u_{i_1} \},u_{i_1} \big)\nonumber\\
&=\underset{i^{\prime}_1\in [|\text{Ind}_{\Wc}|]: i^{\prime}_1 \neq \text{Ind}_{\Wc}^{-1}(i)}{\sum}\negmedspace\negmedspace\negmedspace \text{Card}\big(([|\text{Ind}_{\Wc}|]\setminus \{u^{\prime}_{\text{Ind}_{\Wc}^{-1}(i)}\}) \setminus\{u^{\prime}_1,\ldots,u^{\prime}_{\text{Ind}_{\Wc}^{-1}(i)-1},u^{\prime}_{\text{Ind}_{\Wc}^{-1}(i)+1},\ldots, u^{\prime}_{i_1} \},u^{\prime}_{i_1} \big). \label{eq:case 2 other terms cancellation}
\end{align} 

From~\eqref{eq:case 2 num k}-\eqref{eq:case 2 other terms cancellation}, and the fact that $(-1)^{2 a}=(-1)^0$ for any integer $a$, we have 
\begin{align}
&(-1)^{\underset{i_1\in [|\text{Ind}_{\Wc}|]}{\sum}\negmedspace\negmedspace\negmedspace \text{Card}([|\text{Ind}_{\Wc}|]\setminus\{u_1,\ldots,u_{i_1} \},u_{i_1})+  \underset{i^{\prime}_1\in [|\text{Ind}_{\Wc}|]}{\sum}\negmedspace\negmedspace\negmedspace \text{Card}([|\text{Ind}_{\Wc}|]\setminus\{u^{\prime}_1,\ldots,u^{\prime}_{i^{\prime}_1} \},u^{\prime}_{i^{\prime}_1})} \nonumber\\
&=(-1)^{\text{Card}([|\text{Ind}_{\Wc}|], u_{\text{Ind}_{\Wc}^{-1}(i)})+\text{Card}([|\text{Ind}_{\Wc}|], u^{\prime}_{\text{Ind}_{\Wc}^{-1}(i)}) }.\label{eq:case 2 only one term left}
\end{align}
     Without loss of generality, we assume     $k<k^{\prime}$.   
     Recall that $ \overline{\text{Ind}}_{\Wc \cup \{\Ac(k)\}}(u_{ \text{Ind}_{\Wc}^{-1}(i) })=k^{\prime} $. By the definition  in~\eqref{eq:def of ind bar -1}, 
     we can see that in $\overline{\text{Ind}}_{\Wc}$, there are $\overline{\text{Ind}}^{-1}_{\Wc}(k^{\prime} ) -1$ elements smaller than $k^{\prime}$. By the assumption, $k<k^{\prime}$. Hence, in   $\overline{\text{Ind}}_{\Wc \cup \{\Ac(k)\} }$, there are $\overline{\text{Ind}}^{-1}_{\Wc}(k^{\prime} ) -2$ elements smaller than $k^{\prime}$. 
     In other words, 
     \begin{align}
u_{\text{Ind}_{\Wc}^{-1}(i)}= \overline{\text{Ind}}^{-1}_{\Wc}(k^{\prime} ) -1,
     \end{align}
    which leads to 
     \begin{align}
     \text{Card}(\{u_1,\ldots,u_{|\text{Ind}_{\Wc}|}\}, u_{\text{Ind}_{\Wc}^{-1}(i)})=\overline{\text{Ind}}^{-1}_{\Wc}(k^{\prime} ) -2.\label{eq:case 2 u}
     \end{align}
     Similarly, recall that  $ \overline{\text{Ind}}_{\Wc \cup \{\Ac(k^{\prime})\}}(u^{\prime}_{ \text{Ind}_{\Wc}^{-1}(i) })=k $. In $\overline{\text{Ind}}_{\Wc}$, there are $\overline{\text{Ind}}^{-1}_{\Wc}(k) -1$ elements smaller than $k$. By the assumption, $k<k^{\prime}$. Hence, in   $\overline{\text{Ind}}_{\Wc \cup \{\Ac(k^{\prime})\} }$, there are $\overline{\text{Ind}}^{-1}_{\Wc}(k ) -1$ elements smaller than $k $. In other words, 
\begin{align}
 u^{\prime}_{\text{Ind}_{\Wc}^{-1}(i)}=\overline{\text{Ind}}^{-1}_{\Wc}(k ), 
\end{align}     
     which leads to
     \begin{align}
     \text{Card}(\{u^{\prime}_1,\ldots,u^{\prime}_{|\text{Ind}_{\Wc}|}\}, u^{\prime}_{\text{Ind}_{\Wc}^{-1}(i)})=\overline{\text{Ind}}^{-1}_{\Wc}(k) -1.\label{eq:case 2 u prime}
     \end{align}
     We take~\eqref{eq:case 2 u} and~\eqref{eq:case 2 u prime} into~\eqref{eq:case 2 only one term left} to obtain,
     \begin{subequations}
     \begin{align}
     &(-1)^{\underset{i_1\in [|\text{Ind}_{\Wc}|]}{\sum}\negmedspace\negmedspace\negmedspace \text{Card}([|\text{Ind}_{\Wc}|]\setminus\{u_1,\ldots,u_{i_1} \},u_{i_1})+  \underset{i^{\prime}_1\in [|\text{Ind}_{\Wc}|]}{\sum}\negmedspace\negmedspace\negmedspace \text{Card}([|\text{Ind}_{\Wc}|]\setminus\{u^{\prime}_1,\ldots,u^{\prime}_{i^{\prime}_1} \},u^{\prime}_{i^{\prime}_1})} \nonumber\\
     &=(-1)^{\overline{\text{Ind}}^{-1}_{\Wc}(k^{\prime} ) -2+\overline{\text{Ind}}^{-1}_{\Wc}(k) -1}\\
     &=(-1)^{\overline{\text{Ind}}^{-1}_{\Wc}(k^{\prime} ) +  \overline{\text{Ind}}^{-1}_{\Wc}(k) +1}. \label{eq:case 2 after plus u}
     \end{align}
     \end{subequations}
Finally, we have 
\begin{subequations}
\begin{align}
&(-1)^{ -\overline{\text{Ind}}^{-1}_{\Wc}(k ) - \overline{\text{Ind}}^{-1}_{\Wc}(k^{\prime} )  +\negmedspace\negmedspace\negmedspace \underset{i_1\in [|\text{Ind}_{\Wc}|]}{\sum}\negmedspace\negmedspace\negmedspace \text{Card}([|\text{Ind}_{\Wc}|]\setminus\{u_1,\ldots,u_{i_1} \},u_{i_1})+\negmedspace\negmedspace\negmedspace \underset{i^{\prime}_1\in [|\text{Ind}_{\Wc}|]}{\sum}\negmedspace\negmedspace\negmedspace \text{Card}([|\text{Ind}_{\Wc}|]\setminus\{u^{\prime}_1,\ldots,u^{\prime}_{i^{\prime}_1} \},u^{\prime}_{i^{\prime}_1}) } \nonumber\\
&=(-1)^{ -\overline{\text{Ind}}^{-1}_{\Wc}(k ) - \overline{\text{Ind}}^{-1}_{\Wc}(k^{\prime} ) +\overline{\text{Ind}}^{-1}_{\Wc}(k^{\prime} ) +  \overline{\text{Ind}}^{-1}_{\Wc}(k) +1}\\
&=-1,
\end{align}  
  \end{subequations}
  which proves~\eqref{eq:case 2 equivalent to prove}.
  
\section{Proof of~\eqref{eq:case 3 cancel two term}}
\label{sec:proof of eq:case 3 cancel two term}  
  To prove~\eqref{eq:case 3 cancel two term}, it is equivalent to prove 
  \begin{align}
  (-1)^{1+ \text{Ind}_{\Wc \cup\{\Lc(i)\} }^{-1}(i)  - \overline{\text{Ind}}^{-1}_{\Wc}(\widetilde{k}  )+\negmedspace\negmedspace\negmedspace \underset{i_1\in [|\text{Ind}_{\Wc }|+1]}{\sum}\negmedspace\negmedspace\negmedspace  \text{Card}([|\text{Ind}_{\Wc }|+1]\setminus\{u_1,\ldots,u_{i_1} \},u_{i_1})+ \negmedspace\negmedspace\negmedspace \underset{\widetilde{i}_1\in [|\text{Ind}_{\Wc}|]}{\sum}\negmedspace\negmedspace\negmedspace \text{Card}([|\text{Ind}_{\Wc}|]\setminus\{\widetilde{u}_1,\ldots,\widetilde{u}_{\widetilde{i}_1} \},\widetilde{u}_{\widetilde{i}_1}) }=-1.
  \label{eq:case 3 equivalent to prove}
\end{align}   
  
 Let us focus on $\underset{i_1\in [|\text{Ind}_{\Wc }|+1]}{\sum}\negmedspace\negmedspace\negmedspace  \text{Card}([|\text{Ind}_{\Wc }|+1]\setminus\{u_1,\ldots,u_{i_1} \},u_{i_1})$. By the definition of the function $\text{Card}(\cdot)$ in~\eqref{eq:def of NUM}, we have
 \begin{subequations}
 \begin{align}
& \underset{i_1\in [|\text{Ind}_{\Wc }|+1]}{\sum}\negmedspace\negmedspace\negmedspace  \text{Card}([|\text{Ind}_{\Wc }|+1]\setminus\{u_1,\ldots,u_{i_1} \},u_{i_1}) \nonumber\\
 &= \negmedspace\negmedspace\negmedspace  \underset{\substack{ i_1\in [|\text{Ind}_{\Wc }|+1]:\\ i_1 \neq \text{Ind}_{\Wc \cup\{\Lc(i)\} }^{-1}(i) }}{\sum} \negmedspace\negmedspace\negmedspace \negmedspace\negmedspace\negmedspace  \text{Card}\big(([|\text{Ind}_{\Wc }|+1]\setminus \{u_{\text{Ind}_{\Wc \cup\{\Lc(i)\} }^{-1}(i)} \}) \setminus\{u_1,\ldots,u_{\text{Ind}_{\Wc \cup\{\Lc(i)\} }^{-1}(i)-1}, u_{\text{Ind}_{\Wc \cup\{\Lc(i)\} }^{-1}(i)+1},\ldots, u_{i_1} \},u_{i_1} \big) \nonumber\\
 &+|\{i_2 \in [\text{Ind}_{\Wc \cup\{\Lc(i)\} }^{-1}(i)-1]:u_{\text{Ind}_{\Wc \cup\{\Lc(i)\} }^{-1}(i)}< u_{i_2} \}| \nonumber\\ & + |\{i_3 \in [\text{Ind}_{\Wc \cup\{\Lc(i)\} }^{-1}(i)+1:|\text{Ind}_{\Wc }|+1 ]:u_{i_3}< u_{|\text{Ind}_{\Wc }|+1} \}|\\
 &= \negmedspace\negmedspace\negmedspace  \underset{\substack{ i_1\in [|\text{Ind}_{\Wc }|+1]:\\ i_1 \neq \text{Ind}_{\Wc \cup\{\Lc(i)\} }^{-1}(i) }}{\sum} \negmedspace\negmedspace\negmedspace \negmedspace\negmedspace\negmedspace  \text{Card}\big(([|\text{Ind}_{\Wc }|+1]\setminus \{u_{\text{Ind}_{\Wc \cup\{\Lc(i)\} }^{-1}(i)} \}) \setminus\{u_1,\ldots,u_{\text{Ind}_{\Wc \cup\{\Lc(i)\} }^{-1}(i)-1}, u_{\text{Ind}_{\Wc \cup\{\Lc(i)\} }^{-1}(i)+1},\ldots, u_{i_1} \},u_{i_1} \big) \nonumber\\
 &+ \text{Ind}_{\Wc \cup\{\Lc(i)\} }^{-1}(i)-1- |\{i_2 \in [\text{Ind}_{\Wc \cup\{\Lc(i)\} }^{-1}(i)-1]: u_{i_2}< u_{\text{Ind}_{\Wc \cup\{\Lc(i)\} }^{-1}(i)} \}| \nonumber\\ 
 & + \text{Card}([|\text{Ind}_{\Wc }|+1] ,u_{\text{Ind}_{\Wc \cup\{\Lc(i)\} }^{-1}(i)} )-|\{i_2 \in [\text{Ind}_{\Wc \cup\{\Lc(i)\} }^{-1}(i)-1]:u_{i_2}< u_{\text{Ind}_{\Wc \cup\{\Lc(i)\} }^{-1}(i)}  \}|
 \label{eq:case 3 transform leader}
 \end{align}
    \end{subequations}
  From the construction of $\widetilde{\uv}$ in~\eqref{eq:u tilde}, we have 
  \begin{align}
&\underset{\substack{ i_1\in [|\text{Ind}_{\Wc }|+1]:\\ i_1 \neq \text{Ind}_{\Wc \cup\{\Lc(i)\} }^{-1}(i) }}{\sum} \negmedspace\negmedspace\negmedspace \negmedspace\negmedspace\negmedspace  \text{Card}\big(([|\text{Ind}_{\Wc }|+1]\setminus \{u_{\text{Ind}_{\Wc \cup\{\Lc(i)\} }^{-1}(i)} \}) \setminus\{u_1,\ldots,u_{\text{Ind}_{\Wc \cup\{\Lc(i)\} }^{-1}(i)-1}, u_{\text{Ind}_{\Wc \cup\{\Lc(i)\} }^{-1}(i)+1},\ldots, u_{i_1} \},u_{i_1} \big) \nonumber\\
 &= \underset{\widetilde{i}_1\in [|\text{Ind}_{\Wc}|]}{\sum}\negmedspace\negmedspace\negmedspace \text{Card}([|\text{Ind}_{\Wc}|]\setminus\{\widetilde{u}_1,\ldots,\widetilde{u}_{\widetilde{i}_1} \},\widetilde{u}_{\widetilde{i}_1}).\label{eq:case 3 after transforming}
  \end{align}
 From~\eqref{eq:case 3 transform leader} and~\eqref{eq:case 3 after transforming}, and the fact that $(-1)^{2 a}=(-1)^0$ for any integer $a$, we have 
   \begin{subequations}
 \begin{align}
 & (-1)^{1+ \text{Ind}_{\Wc \cup\{\Lc(i)\} }^{-1}(i)  - \overline{\text{Ind}}^{-1}_{\Wc}(\widetilde{k}  )+\negmedspace\negmedspace\negmedspace \underset{i_1\in [|\text{Ind}_{\Wc }|+1]}{\sum}\negmedspace\negmedspace\negmedspace  \text{Card}([|\text{Ind}_{\Wc }|+1]\setminus\{u_1,\ldots,u_{i_1} \},u_{i_1})+ \negmedspace\negmedspace\negmedspace \underset{\widetilde{i}_1\in [|\text{Ind}_{\Wc}|]}{\sum}\negmedspace\negmedspace\negmedspace \text{Card}([|\text{Ind}_{\Wc}|]\setminus\{\widetilde{u}_1,\ldots,\widetilde{u}_{\widetilde{i}_1} \},\widetilde{u}_{\widetilde{i}_1}) }\nonumber\\
 &=(-1)^{ - \overline{\text{Ind}}^{-1}_{\Wc}(\widetilde{k}  )+\text{Card}([|\text{Ind}_{\Wc }|+1],u_{\text{Ind}_{\Wc \cup\{\Lc(i)\} }^{-1}(i)} )}\\
&=-1 ,\label{eq:last step case 3}
 \end{align}
    \end{subequations}
where~\eqref{eq:last step case 3} comes from that 
   $\widetilde{k} :=\overline{\text{Ind}}_{\Wc  }(u_{\text{Ind}_{\Wc \cup\{\Lc(i)\} }^{-1}(i) })$, and thus $u_{\text{Ind}_{\Wc \cup\{\Lc(i)\} }^{-1}(i)} = \overline{\text{Ind}}^{-1}_{\Wc}(\widetilde{k})$.

\bibliographystyle{IEEEtran}
\bibliography{IEEEabrv,IEEEexample}

\end{document}